\documentclass[aps,prd,nofootinbib,preprintnumbers,superscriptaddress,showpacs,showkeys,floatfix,amsmath,amssymb,amsfonts,twocolumn]{revtex4-2}
\usepackage{graphicx}
\usepackage{bm}
\usepackage{array}
\usepackage{multirow}
\usepackage{amsmath} 
\usepackage{amssymb}
\usepackage{amsfonts}
\usepackage{float}
\usepackage{dsfont}  
\usepackage{slashed}  
\usepackage{booktabs}
\usepackage{multirow} 
\usepackage{makecell}
\usepackage{subfigure}
\usepackage[sort&compress]{natbib}
\usepackage{ulem}
\usepackage{empheq}
\usepackage{color}
\usepackage[svgnames,x11names,table]{xcolor}
\usepackage[colorlinks=true,linkcolor=MediumBlue,citecolor=MediumBlue,urlcolor=MediumBlue]{hyperref}
\usepackage[justification=raggedright,singlelinecheck=false]{caption}
\usepackage{subcaption}
\usepackage{appendix}

\newcommand{\pvec}{\mathbf{p}}
\newcommand{\qvec}{\mathbf{q}}

\begin{document}

\title{Higher Mellin moments of the unpolarized PDF of the pion and the kaon from lattice QCD}

\author{Constantia Alexandrou} \email{alexand@ucy.ac.cy}\affiliation{Department of Physics, University of Cyprus, P.O. Box 20537, 1678 Nicosia, Cyprus}\affiliation{Computation-based Science and Technology Research Center, The Cyprus Institute, 20 Kavafi Str., Nicosia 2121, Cyprus}

\author{Simone Bacchio} \affiliation{Computation-based Science and Technology Research Center, The Cyprus Institute, 20 Kavafi Street, Nicosia 2121, Cyprus}

\author{Priyajit Jana}
\affiliation{Department of Physics, University of Cyprus, P.O. Box 20537, 1678 Nicosia, Cyprus}
\affiliation{Computation-based Science and Technology Research Center, The Cyprus Institute, 20 Kavafi Str., Nicosia 2121, Cyprus}

\author{Marcus Petschlies}
\affiliation{Helmholtz-Institut für Strahlen- und Kernphysik, University of Bonn, Germany}
\affiliation{Bethe Center for Theoretical Physics, University of Bonn, Germany}

\author{Luis Alberto Rodriguez Chacon}
\affiliation{Computation-based Science and Technology Research Center, The Cyprus Institute, 20 Kavafi Str., Nicosia 2121, Cyprus}
\affiliation{Department of Physics and Earth Science, University of Ferrara and INFN, Via Saragat 1, 44122 Ferrara, Italy}

\author{Gregoris Spanoudes}
\affiliation{Department of Physics, University of Cyprus, P.O. Box 20537, 1678 Nicosia, Cyprus}

\author{Fernanda Steffens}
\affiliation{Helmholtz-Institut für Strahlen- und Kernphysik, University of Bonn, Germany}
\affiliation{Bethe Center for Theoretical Physics, University of Bonn, Germany}

\author{Carsten Urbach}
\affiliation{Helmholtz-Institut für Strahlen- und Kernphysik, University of Bonn, Germany}
\affiliation{Bethe Center for Theoretical Physics, University of Bonn, Germany}

\author{Urs Wenger}
\affiliation{Institute for Theoretical Physics, Albert Einstein Center for Fundamental Physics, University of Bern, Switzerland}

\begin{abstract}
We present results on the Mellin moments of the unpolarized parton distribution function (PDF) of the  pion and kaon up to the fourth order. The computation is done using one $N_f=2+1+1$ gauge ensemble of twisted mass fermions with quark masses tuned to  approximately their physical values.   We  reconstruct the valence pion and kaon PDFs using the connected contributions to the three Mellin moments. We  compare our results on the Mellin moments and the reconstructed PDFs with other lattice QCD and phenomenological determinations.
\end{abstract}

\maketitle

\section{Introduction}

Quantum chromodynamics is the theory of the strong interaction, one of the three forces unified in the standard model of particle physics. 
QCD gives rise to the spectrum of hadrons, which represent the physically observable states at low energies, confining quarks and gluons.
Examples for hadrons are protons and neutrons, which dominantly form the matter surrounding us. 

A special role in the zoo of hadronic states is taken by pions and kaons: as part of the lightest pseudoscalar octet they are the lowest mass composite objects to be described by QCD because they are pseudo-Goldstone bosons of spontaneously broken chiral symmetry.
The key difference between pions and kaons is the fact that kaons include valence light and strange quarks, while pions consist of light valence quarks only.
Thus, by comparing them, we have a direct way to study the SU(3) flavor symmetry breaking in QCD.
Beyond the masses of pions and kaons, such a comparison can be performed by studying their partonic structure, information that is encoded in the so-called parton distribution functions (PDFs).
PDFs describe how quarks and gluons (i.e., partons) are distributed in a particular hadronic state depending on Björken's $x$.
For instance, the lowest Mellin moment $\langle x\rangle$ of the unpolarized PDF is interpreted as the average momentum fraction of a parton in a hadron.
Beyond such a comparison and the direct interest in PDFs, they are also important ingredients for experimental analyses.

Experimentally, the structure of the pion and especially the kaon is much less well known than the one of the nucleon. Since pions and kaons are unstable in the standard model, they cannot be used as fixed targets in the same way as protons, and their PDFs can only be extracted indirectly. Existing data for the pion come mainly from Drell-Yan measurements, with additional information from leading-neutron productions analyses~\cite{NA3:1983ejh,E615:1989bda,H1:2010hym}. For the kaon the situation is worse, with experimental data obtained some 40 years ago~\cite{Saclay-CERN-CollegedeFrance-EcolePoly-Orsay:1980fhh}. However, pion and kaon PDFs will be measured by the upcoming AMBER experiment at CERN~\cite{Adams:2018pwt} and at the future Electron-Ion Collider (EIC)~\cite{Aguilar:2019teb}. 
With these experiments in the future, new analyses of the available experimental data are being pursued, in which theoretical results are included.
An example is recent work by JAM~\cite{Barry:2025wjx}, which used lattice-QCD input in their latest global analysis, showing that the information provided is crucial and timely for the determination of parton distribution functions.

Therefore, determining the Mellin moments and the PDFs theoretically will provide significant input for the design of these experiments as well as the interpretation of existing and future results.

On the theoretical side, a variety of model calculations are available, including global QCD fits, Dyson-Schwinger studies, light-front approaches, and related frameworks~\cite{Bednar:2018mtf, Ding:2019qlr, Cui:2020tdf, Hecht:2000xa, Lan:2019rba, Watanabe:2017pvl}. While such studies have provided important insight into the valence structure of the pion and kaon, the spread among such model calculations shows that further first-principles input is still needed, in particular so because often these models are not systematically improvable.

Lattice QCD, on the other hand, provides a systematically improvable, first-principle approach for determining hadron structure. Substantial progress has been made in lattice-QCD studies of meson structure, using different methods such as local operators for Mellin moments~\cite{Alexandrou:2026soz}, approaches aiming at the determination of the $x$-dependence of the PDFs using large-momentum effective theory through quasi distributions~\cite{Ji:2014gla}, and other related approaches~\cite{PhysRevD.96.034025, PhysRevD.96.094503}. 
Recently, novel methods based on the gradient flow~\cite{Shindler:2023xpd} have been proposed to enable the computation of Mellin moments beyond the fourth order, with first studies providing promising results~\cite{z3wr-zk8n}. 
In a previous work by the Extended Twisted Mass Collaboration (ETMC), the first full decomposition of the momentum fractions for the quarks and gluons in the continuum limit was performed for the pion and kaon including disconnected contributions and directly at the physical pion mass~\cite{ExtendedTwistedMass:2024kjf}. 

In this work, we extend our previous work and calculate the third and fourth Mellin moments, $\langle x^2\rangle$ and $\langle x^3\rangle$, respectively, of the pion and kaon valence quark unpolarized PDFs using local operators.
The calculation is performed using one $N_f = 2+1+1$ gauge  ensemble of twisted mass clover-improved fermions with all quark masses tuned to approximately their  physical value.
We discuss the SU$(3)$ symmetry breaking by comparing  ratios of moments of light and strange quarks in the kaon. Furthermore,  with the three moments available, we reconstruct the $x$-dependence of the PDFs by fitting to the standard functional form, which allows us to compare with other determinations of PDFs from phenomenology and lattice QCD.

This paper is organized as follows: in Sec.~\ref{sec:theory}, we describe the theoretical framework and the decompositions to extract the moments. In Sec.~\ref{sec:Analysis_of_lattice_data}, we describe the lattice setup and the  analysis of lattice correlators to extract each moment for the pion and the kaon as well as their renormalization. In Sec.~\ref{sec:results}, we present our results for the moments and discuss SU(3) symmetry breaking. In Sec.~\ref{sec:PDF_reconstruction}, we present a reconstruction of the PDFs using the calculated moments. Section~\ref{sec:comparison_with_other_determinations} includes  comparison of our moments to other lattice QCD, experimental data analysis and other model calculations and compares our reconstructed PDF to experimental data analysis determinations. In Sec.~\ref{sec:conclusions}, we present conclusions and perspectives for future work.

\section{Matrix elements decomposition}
\label{sec:theory}
The momentum fraction $\langle x \rangle_f^M$ is extracted from the matrix element of the traceless energy-momentum tensor (EMT),
\begin{equation}
  \label{eq:x}
  \langle M(\pvec) \,|\, \bar{T}_{f;\mu\nu} \,|\, M(\pvec) \rangle =
  2\langle x\rangle^M_f\left(p_\mu p_\nu - \delta_{\mu\nu}\frac{p^2}{4}\right),
\end{equation}
where $M$ denotes the pion or the kaon meson and $f$  the quark flavor. The EMT in Euclidean spacetime is given by
\begin{equation}
  \bar{T}_{f;\mu\nu}\ =\ -\frac{(i)^{\kappa_{\mu\nu}}}{2}\,\bar q_f\,
  \left(
    \gamma_\mu\stackrel{\leftrightarrow}{D}_\nu + \gamma_\nu \stackrel{\leftrightarrow}{D}_\mu
    - \delta_{\mu\nu}\,\frac{1}{2}\,\gamma_\rho\,\stackrel{\leftrightarrow}{D}_\rho
  \right)q_f\,,
  \label{Tq}
\end{equation}
with $\kappa_{\mu\nu} = \delta_{\mu,4}\,\delta_{\nu,4}$,
and the symmetrized covariant derivative
$ \stackrel{\leftrightarrow}{D_\mu} \, = \, \frac{1}{2}(\stackrel{\rightarrow}{D}_\mu - \stackrel{\leftarrow}{D}_\mu)$. The momentum fraction $\langle x \rangle_f^M$ was computed for the same ensemble as used here in Ref.~\cite{ExtendedTwistedMass:2024kjf} and will be used in this work.

In general, the even unpolarized moments of the quark PDFs are purely valence, while the odd moments have also contributions from sea quarks and antiquarks, according to
\begin{equation}
        \langle x^n \rangle_q  
    = \int_0^1 d{x} x^n \big[ q(x,\mu)  + (-1)^{n+1} \bar q(x,\mu) \big],
\end{equation}
with $\mu$ the renormalization scale, and the valence distributions is defined as $q_v(x,\mu) \equiv q(x,\mu) - \bar q(x,\mu)$.
Thus, while the $\langle x^2\rangle$ moment is given solely by the valence distribution, the odd moments are not. In these cases, we make the assumption that the valence distributions are dominated by the connected parts, and we do not include the disconnected insertions in the computation of both the $\langle x\rangle$ and $\langle x^3 \rangle$ moments.

For the computation of higher moments, we need to compute the matrix elements of operators of higher derivatives. The operator product expansion relates the $n{\rm th}$ moment  
$\langle x^{n-1} \rangle$ to the matrix element of the operator $\langle\mathcal{O}^{\mu_1\,\mu_2, \dots \mu_n}\rangle$,
where $\mu_j,\, j=1,\cdots,n$ are Lorentz indices. Thus, in order to compute the third and the fourth moments  we will need to compute the matrix elements of the operators
\begin{align}
\mathcal{O}^{\{\mu\nu\rho\}}_f
&= \bar q_f\, \gamma^{\{\mu}\, \overset{\leftrightarrow}{D}{}^{\nu}\, \overset{\leftrightarrow}{D}{}^{\rho\}}\, q_f, \nonumber\\
\mathcal{O}^{\{\mu\nu\rho\sigma\}}_f
&= \bar q_f\, \gamma^{\{\mu}\, \overset{\leftrightarrow}{D}{}^{\nu}\, \overset{\leftrightarrow}{D}{}^{\rho}\, \overset{\leftrightarrow}{D}{}^{\sigma\}}\, q_f,
\label{eq:operators}
\end{align}
where the curly brackets denote symmetrization over the indices and subtraction of the trace. In order to avoid mixing, we take all indices to be different, which means that we need a boosted frame. The decomposition of the matrix elements in the forward limit is given by \cite{Alexandrou:2020gxs,Alexandrou:2021mmi}
\begin{align}
    \nonumber \langle M(\pvec )| \mathcal{O}^{\{\mu\nu\rho\}}_f | M(\pvec)\rangle =& 2i\mathcal{K} p^{\{\mu}p^{\nu}p^{\rho\}}  \langle x^2 \rangle^M_f  \\
    \nonumber\langle M(\pvec )| \mathcal{O}^{\{\mu\nu\rho\sigma\}}_f | M(\pvec)\rangle = &-2\mathcal{K}  p^{\{\mu } p^{\nu} p^{\rho} p^{\sigma \} }  \langle x^3 \rangle^M_f  \\
  \label{eq:final_decomp2}
\end{align}
where $\mathcal{K} = 1/2E(\pvec)$ is a kinematic factor. 
We take one index to be  in the temporal direction and the rest in different spatial directions.  Such a choice avoids mixing with lower dimensional operators but requires boosted meson states.
For $\langle x^2 \rangle^M_f$, we use a boosted frame where $\pvec$ has nonzero values in at least two spatial directions, $\pvec = (2\pi/L,2\pi/L,0)$  and permutations and for $\langle x^3 \rangle^M_f$ $\pvec = (2\pi/L,2\pi/L,2\pi/L)$.

We extract the moments from the ratio of three- to two-point functions.
The three-point function in the forward direction is given by
\begin{align}
 C^{\mathcal{O}}_M(t_{\text{s}}, t_{\text{ins},},t_{0}, \pvec) = &\sum_{\textbf{x}_s, \textbf{x}_{\text{ins}}}\langle 0|J_M(t_{\text{s}},\textbf{x}_{\text{s}})\mathcal{O}_f(t_{\text{ins}},\textbf{x}_{\text{ins}}) \nonumber\\ 
&\times\bar{J}_M(t_{\text{0}},\textbf{x}_{\text{0}})|0\rangle e^{-i\pvec\cdot(\textbf{x}_{\text{s}}-\textbf{x}_{0})},
    \label{eq:c3pts_}
\end{align}
where $(t_0,\textbf{x}_0)$, $(t_{\rm ins}$,\textbf{x}$_{\text{ins}}$) and $(t_{\rm s},\textbf{x}_{\text{s}})$ denote the source, insertion and sink coordinates, respectively, and $\mathcal{O}_f$ are the derivative operators of Eq.~\eqref{eq:operators}. Without loss of generality, we take $(t_0,{\bf x}_0)=(0,\bf{0})$

The two-point functions in the boosted frame are given by
\begin{equation}
    C_M(t_\text{s}, \pvec) = \sum_{\textbf{x}}\langle 0|J_M(t_\text{s},\textbf{x})
    \bar{J}_M(0,\textbf{0})|0\rangle e^{-i\pvec\cdot\textbf{x}},
\end{equation}
and the ratio used to cancel overlaps and exponential factors is 
\begin{equation}
  R^{\mathcal{O}}_M(t_\text{s}, t_\text{ins}, \pvec) = 
  \omega_T(t_\text{s},\mathbf{p}) \, \frac{C^{\mathcal{O}}_M(t_{\text{s}}, t_{\text{ins},}, \pvec)}
{C_M(t_{\text{s}}, \pvec)} .
\label{eq:ratio}
\end{equation}
 The factor
\begin{align*}
  \omega_T(t_\text{s},\qvec) &= 1+\exp[ -E(\qvec) \, (T - 2\,t_\text{s} ) ]
\end{align*}
takes into account the periodicity of the lattice.  If $t_\text{s}-t_0$ and $t_{\text{ins}}-t_0$ are large enough, the ratio in Eq.~\eqref{eq:ratio} becomes independent of the time
\begin{equation}
    R^{\mathcal{O}}_M(t_\text{s}, t_\text{ins}, \pvec) \xrightarrow[t_\text{s}-t_{\mathrm{ins}}\to\infty]{t_\text{s}\to\infty} \Pi^{\mathcal{O}}_M(\pvec),
    \label{eq:pi_ratio}
\end{equation}
The relations to the moments then become 
\begin{equation}
    \langle x^2 \rangle^M_f = -\frac{\Pi^{ij}_M(\pvec)}{p^ip^j}, \quad
    \langle x^3 \rangle^M_f = \frac{i \Pi^{1234}_M(\pvec)}{p^1p^2p^3}.
    \label{eq:moments_from_ratios}
\end{equation}
The three-point functions in Eq.~\eqref{eq:c3pts_} generate connected and disconnected contributions. For the higher moments these are expected to be small and they are neglected here.

\section{Lattice setup and Analysis}\label{sec:Analysis_of_lattice_data}
We use a gauge ensemble generated by the ETMC with $N_f = 2 + 1 + 1$ twisted mass clover-improved fermions with quark masses tuned to approximately their physical values.  In Table~\ref{tab:ensemble} we give the parameters of the ensemble labeled as cB211.072.64~\cite{Alexandrou:2018egz}. The statistics used for this work are shown in Table~\ref{tab:stats}.

\begin{table}[th]
  \caption{Parameters of the ETMC gauge ensemble used in this work, where  $a$ is the lattice spacing,  $L$ the spatial and $T$ the temporal lattice size, and
    $M_{\pi^{\pm}}$ and $M_{K^{\pm}}$ the  pion and kaon masses, respectively.
  }
\begin{tabular}{cccccc}
\hline
Name & a [fm] & $L^3\times T$ &  $M_\pi$~[MeV] & $M_{K}$~[MeV] \\
\hline
cB211.072.64 &  $0.0796(1)$ & $64^3\times 128$  & $140.40(22)$  &  $498.41(11)$\\
\hline
\end{tabular}
\label{tab:ensemble}
\end{table}

\begin{table}[t]
\centering
\caption{Statistics used in the calculation of $\langle x^2\rangle$ and $\langle x^3\rangle$. We give the number of configurations $(n_{\rm conf})$, the source sink-separations in the case of the three-point correlators $(t_s/a)$, the number of stochastic sources   per configuration $(n_{\rm sts})$ and the number of momentum boosts $(n_{\rm boost})$. Last column shows the total statistics.}
\label{tab:stats}
\renewcommand{\arraystretch}{1.15}
\begin{tabular*}{\linewidth}{@{\extracolsep{\fill}} c c c c c c @{}}
\toprule
\toprule
& \multicolumn{5}{c}{Statistics} \\
\cmidrule(lr){1-6}
$\langle x^n\rangle$
& $t_s/a$
& $n_{\mathrm{conf}}$
& $n_{\mathrm{sts}}$
& $n_{\mathrm{boost}}$
& \makecell{Total \\ statistics} \\
\midrule

\multicolumn{6}{c}{Three-point correlators} \\
\midrule
$\langle x^2 \rangle$  
& 16, 20, 24
& 786 
& 8 
& 12 
& 75,456 \\

$\langle x^3 \rangle$ 
& 16, 20, 24
& 736
& 8
& 8 
& 47,104 \\

\bottomrule
\bottomrule
\end{tabular*}
\end{table}

We compute statistical  errors using a jackknife analysis. Systematic errors due to the fitting procedure are calculated using model averaging with weights based on the Akaike Information Criterion (AIC)~\cite{Jay:2020jkz}, $\omega_i$ given by
\begin{equation}
    \text{ln}(\omega_i) = -\frac{1}{2} \chi_{i}^2 + N_{\mathrm{dof}, i},
\end{equation}
where $N_\mathrm{dof} = N_{\rm data} - N_{\rm params}$ with $N_{\rm data}$ the number of measurements and $N_{\rm params}$ the number of fit parameters. The probability for each fit is defined as
\begin{equation}
    p_i = \frac{\omega_i}{\sum_i \omega_i}.
\end{equation}
The model-averaged mean and error for an observable $\mathcal{O}$ is then given by
\begin{equation}\begin{split}
    \langle \mathcal{O} \rangle &= \sum_i \mathcal{O}_i p_i, \\
    \sigma_\mathcal{O}^2 &= \sum_i \sigma_{\mathcal{O}_i}^2 p_i + \sum_i \mathcal{O}_{i}^2 p_i - \langle \mathcal{O}\rangle^2,
    \end{split}
\end{equation}
where $\mathcal{O}_i$ and $\sigma_{\mathcal{O}_i}$ are the mean and error, respectively, for each fit.

\subsection{Determination of ground-state energies}

\begin{figure}
    \centering
    \includegraphics[width=\linewidth]{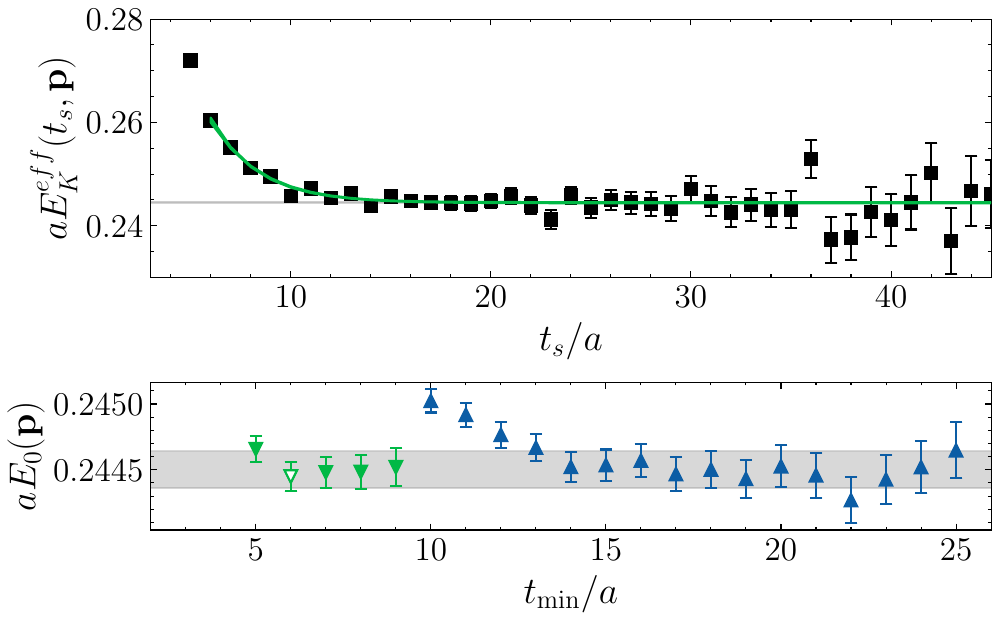}
    \caption{Upper panel: Results on the effective energy extracted using Eq.~\eqref{Eq:Eeff} for the kaon, with $\pvec^2 = 8\pi^2/L^2$ as a function of the time separation $t_\text{s}/a$ in lattice units.  Lower panel: Convergence as one varies the lower fit range $t_{\rm min}/a$.  The values extracted are shown when using one-state fit (blue triangles)  and the two-state fit (green triangles), respectively. The open symbol indicates the highest probability according to the AIC and the gray band is the model-averaged value for the ground-state energy.}
    \label{fig:eff_energy}
\end{figure}

The spectral decomposition of a two-point meson correlation function is given by
\begin{align}
        C^{2pt}(t_\text{s}, \textbf{p}) =& 
        \sum_{n=0}^{\infty} c_n(\textbf{p}) \left (e^{-E_n(\textbf{p})t_\text{s}} + e^{-E_n(\textbf{p})(T - t_\text{s})} \right)\, ,
        \label{eq:c2pt_ans}
\end{align}
taking into account the periodicity of meson two-point functions. 
With $C_M$, we define the effective energy $E_\mathrm{eff}$ as follows
\begin{equation}
    E_{\text{eff}}(t_\text{s}, \pvec) = \text{cosh}^{-1}\left[\frac{C_M(t_{s}-1, \pvec) + C_M(t_\text{s}+1, \pvec)}{2C_M(t_\text{s}, \pvec)}\right]\, .
    \label{Eq:Eeff}
\end{equation}
$E_\mathrm{eff}$ converges to the ground-state energy once excited states have decayed. 
Thus, to determine the ground-state energy level $E_0$ from two-point function data, we perform a one- and two-state analysis of the effective energy $E_\mathrm{eff}(t_\text{s})$. For one-state fits we have one fitting parameter, while for two-state fits we have three fitting parameters, namely the ratio $c_1(\pvec)/c_0(\pvec)$, the ground-state energy $E_0(\pvec)$ and the first excited state energy $E_1(\pvec)$. In Fig.~\ref{fig:eff_energy} we show as an example the effective energy for the kaon with $\pvec^2 = 8\pi^2/L^2$, and the ground-state energy extracted from the fits. As can be seen, the results from the plateau (one-state) fit are compatible with the two-state fit. We take the ground-state energy $E_0(\pvec)$  as determined  from the two-state fit with the highest probability according to the AIC, which is then used as an input parameter with uncertainties in the following.

\subsection{Determination of the moments}

We extract the moments from the ratios  
given in Eq.~\eqref{eq:ratio} as explained in this section. The spectral decomposition of the three-point function keeping terms up to the first excited state is given by
\begin{align}
    \nonumber C^{3pt}(\textbf{p},t_\text{s}, t_\text{ins}) &= A_{00}(\pvec)e^{-E_0(\textbf{p})t_s} \\
    \nonumber &+ A_{01}(\pvec)e^{-E_0(\textbf{p})(t_\text{s}-t_\text{ins})-E_1(\textbf{p}) t_\text{ins}} \\ 
    \nonumber &+ A_{10}(\pvec)e^{-E_1(\textbf{p})(t_\text{s}-t_\text{ins})-E_0(\textbf{p}) t_\text{ins}} \\
              &+ A_{11}(\pvec)e^{-E_1(\textbf{p})t_\text{s}}.
    \label{eq:c3pt_corr}
\end{align}

\begin{figure}[h!]
    \centering
    \includegraphics[width=\columnwidth]{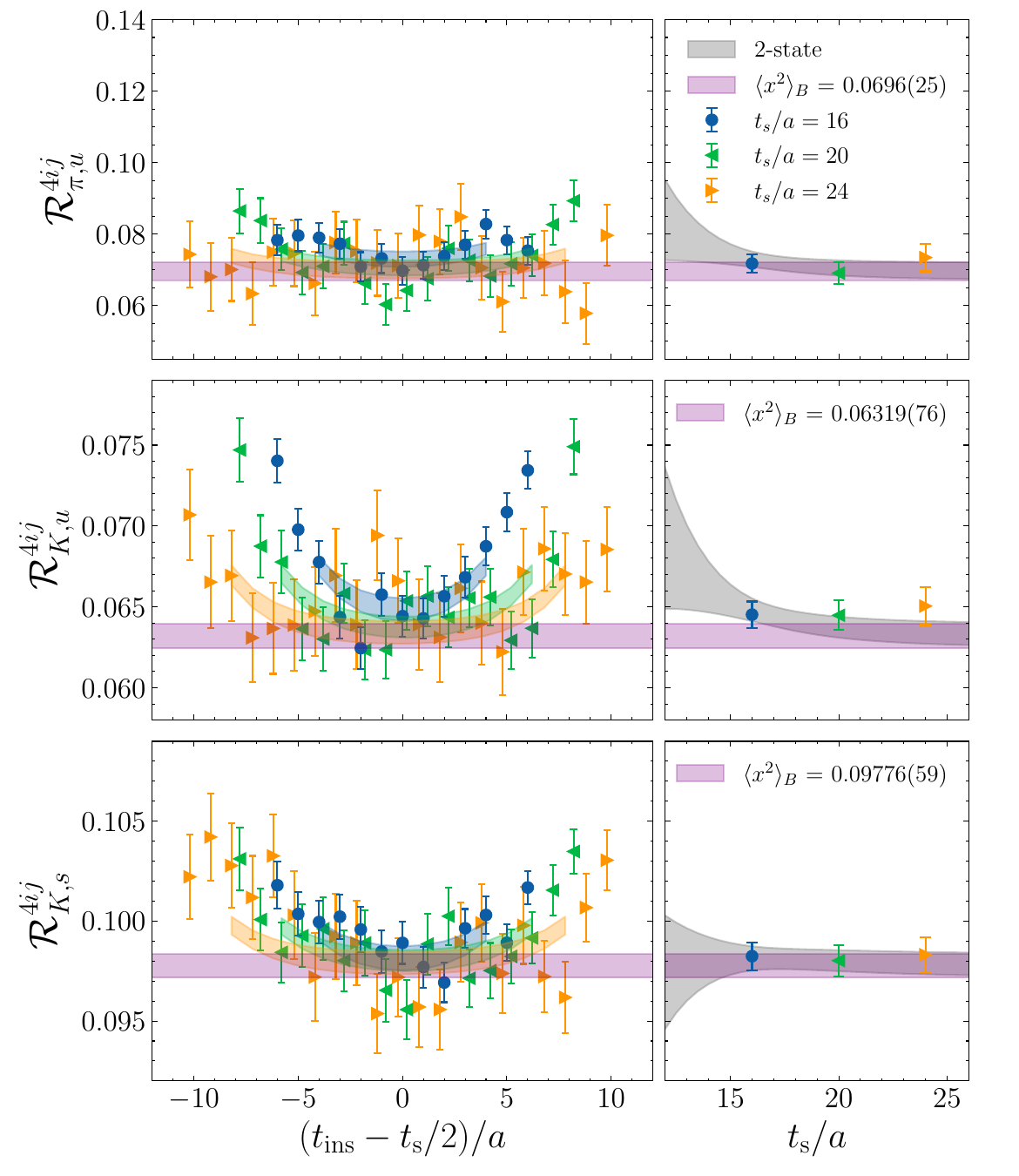}
    \caption{Left column: Results on the bare ratios of three- to two-point functions from which   $\langle x^2 \rangle$ are extracted  for three source-sink time separations $t_s/a = 16, 20, 24$.  The bands are fits using two-state analysis.  From top to bottom, we show $\langle x^2 \rangle_u^\pi$,  $\langle x^2 \rangle_u^K$ and $\langle x^2 \rangle_s^K$. Right column: Results from one-state fit to  each source-sink time separation (blue, green and yellow points). The gray band is the resulting curve when one uses the highest probability parameters determined from the two-state fit analysis. The purple horizontal band across both columns shows the final value determined using model averaging of two-state fits. }
    \label{fig:ratios_x2}
\end{figure}
\begin{figure}[h!]
    \centering
    \includegraphics[width=\columnwidth]{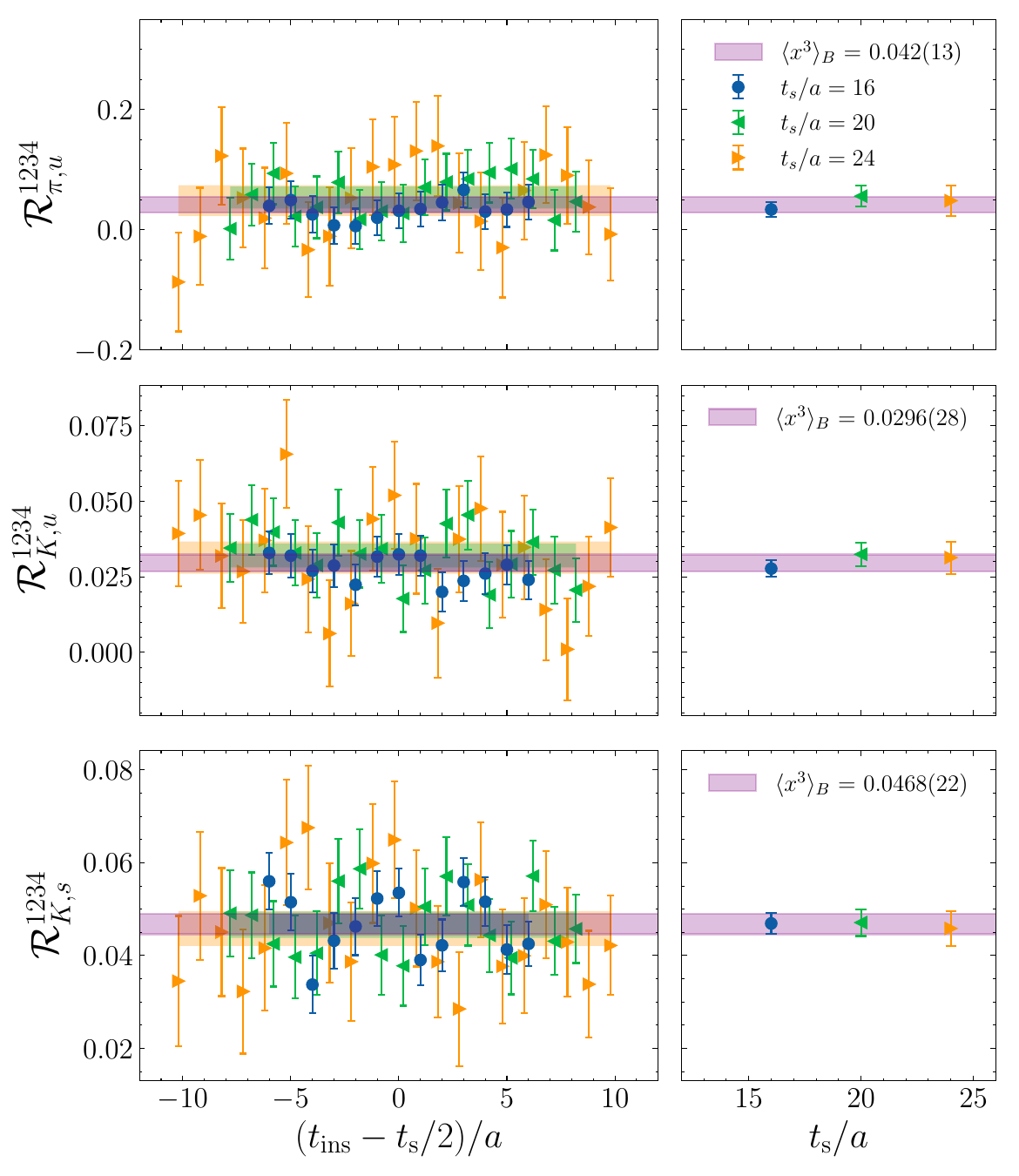}
    \caption{Left column: Results on the bare ratios of three- to two-point functions from which  $\langle x^3 \rangle$ are extracted. Right column: Results from one-state fits to the ratios at each time separation. The purple band is the model average of the result at the three time separations. The rest of the notation is the same as that  in Fig.~\ref{fig:ratios_x2}.} 
    \label{fig:ratios_x3}
\end{figure}

In the case of zero momentum transfer, as is the case considered here, the factors $A_{01}(\pvec)=A_{10}(\pvec)$~\cite{Alexandrou:2021mmi}. We fit simultaneously the effective energy and the ratios between three- and two-point functions for several combinations of $t_s$ and $t_{\rm ins}$ using the expressions \eqref{eq:c2pt_ans} and \eqref{eq:c3pt_corr} up to the first excited state. The ground-state energy $E_0(\pvec)$ is fixed to the highest probability values extracted from the two-state analysis of the pion and kaon effective energies, and is therefore not included among the fit parameters. The fitting parameters are the amplitudes in Eq.~\eqref{eq:c3pt_corr}, the first excited-state energy $E_1(\pvec)$, and the overlap-factor ratio $c_1(\pvec)/c_0(\pvec)$, with the latter two shared by the effective-energy and ratio. From Eqs.~\eqref{eq:pi_ratio}, \eqref{eq:c3pt_corr} and \eqref{eq:c2pt_ans} it can be deduced that the ratio, in the large-time limit,  is related to the fitted parameters 
\begin{equation}
\Pi^{\mathcal{O}}_M = \frac{A_{00}}{{c_0}}.
\end{equation}
We show in Fig.~\ref{fig:ratios_x2} the bare ratios
of the three-point function involving the operator $\mathcal{O}^{\{\mu\nu\rho\}}_f $ used for the extraction of $\langle x^2 \rangle_B$, where the $B$ subindex refers to the bare quantity. We also show the resulting values of $\langle x^2 \rangle$  that converges for the different fit ranges used. 

We show in Fig.~\ref{fig:ratios_x3} the bare ratios used in the extraction of
$\langle x^3 \rangle$. They show fast convergence with no clear 
signal of excited states contamination, within the current errors.
Given the lack of detectable excited states, we perform a simultaneous
one-state fit using the three source-sink time separations to extract the moments. 

\subsection{Renormalization}
\label{ssec:renormalization}

The matrix elements of $\mathcal{O}^{\{\mu\nu\rho\}}_f$ and 
$\mathcal{O}^{\{\mu\nu\rho\sigma\}}_f$ are renormalized nonperturbatively 
in the RI$'$-MOM scheme~\cite{Martinelli:1994ty}, followed by 
a perturbative conversion to the $\overline{\rm MS}$ scheme 
at the scale $\mu = 2$ GeV. 
In general, operators with two or more covariant derivatives exhibit 
a nontrivial renormalization pattern on the lattice including mixing 
with operators of equal or lower dimension as allowed by hypercubic symmetry~\cite{Gockeler:1996mu}. 
For the cases of two- and three-derivative operators considered in 
this work, such mixing is absent when their Lorentz indices $(\mu,\nu,\rho)$ 
and $(\mu,\nu,\rho,\sigma)$ are all different. This choice greatly 
simplifies the renormalization procedure, as it avoids the potential 
need for delicate subtractions associated with power-divergent mixing. 
Restricting ourselves to this choice of indices, we determine the 
multiplicative renormalization functions of the two operators, denoted  by $Z_{\rm VDD}$ and 
$Z_{\rm VDDD}$. These are computed nonperturbatively following the 
procedure outlined in, e.g., Refs.~\cite{ExtendedTwistedMass:2024kjf,Alexandrou:2024ozj,Alexandrou:2025vto}. Note that flavor-singlet three-derivative operators can, in 
principle, mix with gluon two-derivative operators in both continuum and 
lattice regularizations. However, this mixing is expected to be much weaker than that of the flavor-singlet one-derivative operator studied by us in Ref.~\cite{ExtendedTwistedMass:2024kjf}. Since disconnected matrix elements are not considered in this work for the two- and three-derivative operators, we restrict the analysis to multiplicative nonsinglet renormalization functions.

The RI$'$-MOM scheme is defined on amputated vertex functions of the 
operators under study with external off-shell quark states in Landau gauge, as follows:
\begin{equation}
    \Lambda_{\mathcal{O}} (p) = \frac{a^{12}}{V} \sum_{x,y,z} e^{-i p (x-y)} \langle q (x) \mathcal{O} (z) \bar{q} (y) \rangle_{\rm amp.}, 
\end{equation}
where $\mathcal{O} \in (\mathcal{O}^{\{\mu\nu\rho\}}, \mathcal{O}^{\{\mu\nu\rho\sigma\}})$. In the continuum limit, $\Lambda_{\mathcal{O}} (p)$ is decomposed into two independent structures allowed by Lorentz symmetry~\cite{Gracey:2006zr}, as follows:
\begin{eqnarray}
      \Lambda_{\mathcal{O}^{\{\mu\nu\rho\}}} (p) &=& - \gamma^{\{\mu} p^\nu p^{\rho \}} \ \Sigma^{\rm VDD}_1 (p^2) - \nonumber \\
      && \frac{\slashed{p}}{p^2} p^{\{\mu} p^\nu p^{\rho \}} \ \Sigma^{\rm VDD}_2 (p^2), 
      \label{LVDD} \\
      \Lambda_{\mathcal{O}^{\{\mu\nu\rho\sigma\}}} (p) &=& - i \gamma^{\{\mu} p^\nu p^{\rho} p^{\sigma \}} \ \Sigma^{\rm VDDD}_1 (p^2) - \nonumber \\
      && i \frac{\slashed{p}}{p^2} p^{\{\mu} p^\nu p^\rho p^{\sigma \}} \ \Sigma^{\rm VDDD}_2 (p^2), \label{LVDDD}
    \end{eqnarray}
    where $\Sigma^{\rm X}_1 (p^2) = 1 + \mathcal{O} (\alpha_s)$, and $\Sigma^{\rm X}_2 (p^2) = \mathcal{O} (\alpha_s)$, for X $=$ VDD, VDDD. In continuum regularizations, the renormalization conditions are typically defined in terms of the first form factor, $\Sigma^{\rm X}_1(p^2)$~\cite{Gracey:2006zr}, which can be isolated by applying suitable projectors to the vertex functions. On the lattice, however, the available projectors should be restricted to operators whose Lorentz indices are all different. This constraint complicates the construction of projectors that cleanly isolate $\Sigma^{\rm X}_1$. In particular, for the three-derivative operator, the two tensor structures become nearly indistinguishable for momenta close to the body-diagonal direction. In this case, one may instead retain a single structure, namely the first one appearing in Eq.~\eqref{LVDDD} that is present at tree level, and use its inverse as a projector. Based on these considerations, we impose the following conditions at $p^2 = \mu_0^2$ to extract $Z_{\rm VDD}$ and $Z_{\rm VDDD}$:\footnote{{In principle, one can construct a projector that isolates $\Sigma_1^{\rm VDDD}$, namely $P^{\mu\nu\rho\sigma} = 4i (\tilde{p}^2 \sum_\tau \gamma_\tau/\tilde{p}_\tau - 4 \slashed{\tilde{p}}) [\tilde{p}^\mu \tilde{p}^\nu \tilde{p}^\rho \tilde{p}^\sigma (\sum_\tau \tilde{p}^2/\tilde{p}_\tau^2 - 16)]^{-1}$. However, this projector becomes impractical for momenta close to the body-diagonal direction, where its denominator approaches zero.}}
    \begin{eqnarray}
      {(Z_q^{{\rm RI}'})}^{-1} Z_{\rm VDD}^{{\rm RI}'} \frac{1}{12} \sum_{\mu<\nu<\rho} {\rm Tr} \left[ \Lambda_{\mathcal{O}^{\{\mu\nu\rho\}}} P^{\mu\nu\rho} \right] &=& 1, \label{ZVDD} \\
      {(Z_q^{{\rm RI}'})}^{-1} Z_{\rm VDDD}^{{\rm RI}'} \frac{1}{12} \sum_{\mu<\nu<\rho<\sigma} {\rm Tr} \left[ \Lambda_{\mathcal{O}^{\{\mu\nu\rho\sigma\}}} P^{\mu\nu\rho\sigma} \right] &=& 1, \qquad \label{ZVDDD}
    \end{eqnarray}
where
\begin{eqnarray}
P^{\mu\nu\rho} &=& \frac{3 \slashed{\tilde{p}} - 4( \slashed{\tilde{p}}^\mu + \slashed{\tilde{p}}^\nu + \slashed{\tilde{p}}^\rho)}{4 \tilde{p}^\mu \tilde{p}^\nu \tilde{p}^\rho}, \\
P^{\mu\nu\rho\sigma} &=& \frac{4 i \sum_\tau \gamma^\tau / \tilde{p}^\tau}{\tilde{p}^\mu \tilde{p}^\nu \tilde{p}^\rho \tilde{p}^\sigma \sum_\tau 1/ (\tilde{p}^\tau)^2},
\end{eqnarray}
and $\tilde{p}^\mu \equiv \sin(a p^\mu)$, $\slashed{\tilde{p}}^\mu \equiv \gamma^\mu \tilde{p}^\mu$ (no sum), $\slashed{\tilde{p}} \equiv \sum_\mu \slashed{\tilde{p}}^\mu$. $\tilde{p}^\mu$, $\tilde{p}^\nu$, $\tilde{p}^\rho$, and $\tilde{p}^\sigma$ are all strictly nonzero. $\mu_0$ represents the RI$'$-MOM scale. $Z_q^{{\rm RI}'}$ is the renormalization factor of the quark field defined by~\cite{ExtendedTwistedMass:2021gbo}:
\begin{equation}
     Z_q^{{\rm RI}'} = \frac{1}{12} \sum_{\mu} {\rm Tr} \left[S^{-1} (p) \cdot \frac{-i \ \gamma^\mu}{4 {\tilde{p}}^\mu}\right] \Big|_{p^2 = \mu_0^2},
\end{equation}
where $S(p)$ is the quark propagator in the momentum space.  
 \begin{figure}[h!]
\centering
\includegraphics[width=\columnwidth]{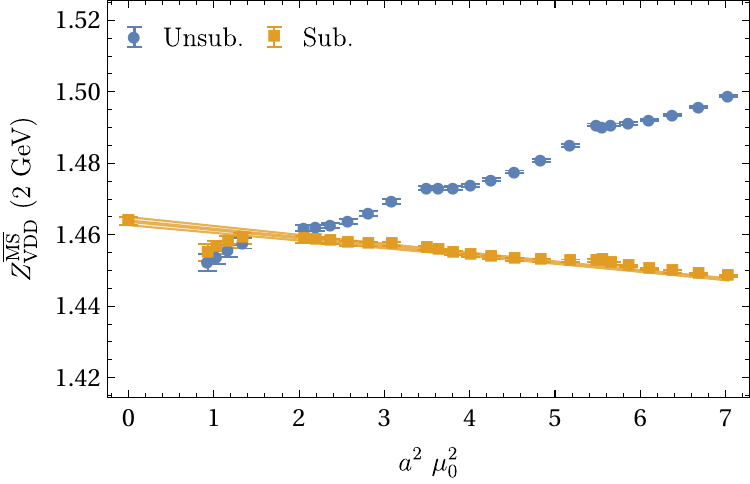} \\
\includegraphics[width=\columnwidth]{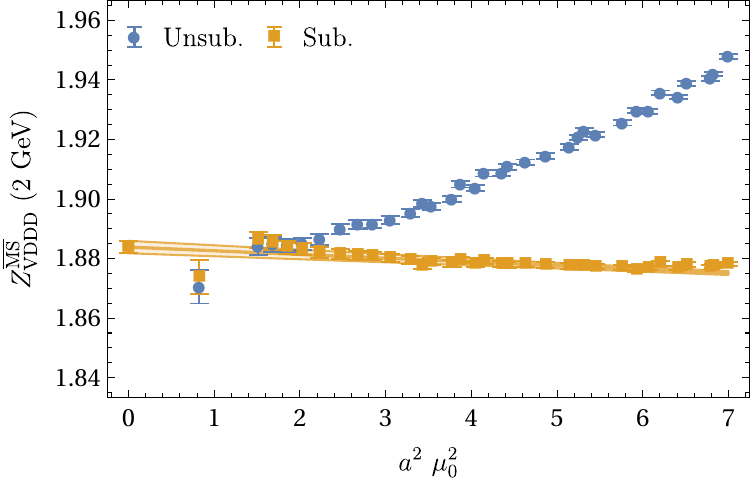}
    \caption{$Z_{\rm VDD}^{\overline{\rm MS}}$ and $Z_{\rm VDDD}^{\overline{\rm MS}}$ as a function of $a^2 \mu_0^2$ at the reference scale of 2 GeV. The data are given with (Sub.) and without (Unsub.) subtracting one-loop artifacts. A linear fit $c_0 + c_1 a^2 \mu_0^2$ is employed in the subtracted data in the ranges $a^2 \mu_0^2 \in [2,5]$ and $a^2 \mu_0^2 \in [2,6]$ for $Z_{\rm VDD}$ and $Z_{\rm VDDD}$, respectively. The extrapolated values $c_0$ are given at $a^2 \mu_0^2 = 0$.}
    \label{fig:Zfactors}
\end{figure}

\begin{figure}[ht!]
\centering
\includegraphics[width=\columnwidth]{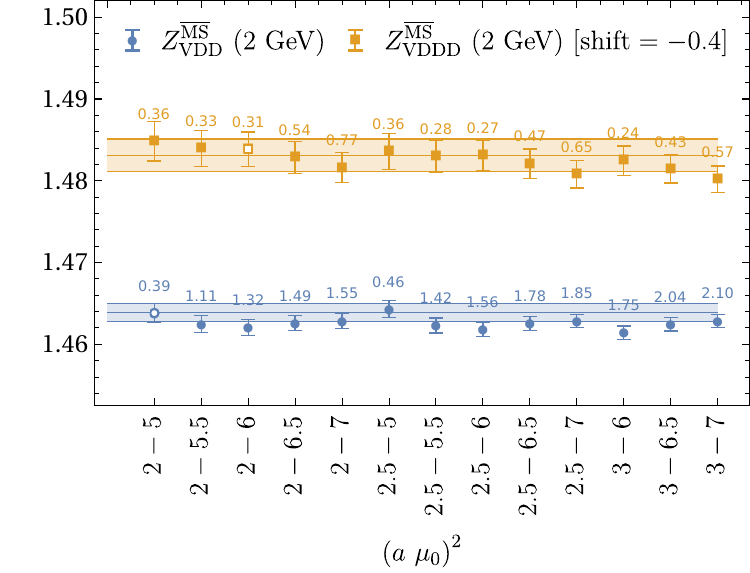} 
    \caption{Extrapolated values of $Z_{\rm VDD}^{\overline{\rm MS}}$ and $Z_{\rm VDDD}^{\overline{\rm MS}}$ at $a^2 \mu_0^2 = 0$ from momentum fits over multiple fit ranges together with the AIC-averaged values (bands). The reduced $\chi^2$ of each fit is given by the label above the corresponding data point. $Z_{\rm VDDD}^{\overline{\rm MS}}$ is shifted by -0.4. The open symbols denote the fits with the highest AIC weights.}
    \label{fig:AIC}
\end{figure}
The vertex functions $\Lambda_{\mathcal{O}} (p)$ and quark propagators $S(p)$ are calculated using Landau gauge fixed momentum sources~\cite{Gockeler:1998ye}, which leads to high statistical accuracy using only 30 configurations. We employ four ensembles specifically simulated for our renormalization program (see Ref.~\cite{Alexandrou:2024ozj}), which features four mass-degenerate quarks ($N_f = 4$) at the same $\beta$ value as the physical-point ensemble used in our analysis of the matrix elements. The ensembles correspond to different twisted-mass parameters $\mu_{\rm sea}$, or equivalently pion masses, and are used in order to take the chiral limit. 
The dependence on $\mu_{\rm sea}$ is found to be mild, consistent with previous studies of our group at a coarser lattice spacing~\cite{Alexandrou:2020gxs,Alexandrou:2021mmi}. We remove this dependence by performing a linear fit in $\mu_{\rm sea}$. 

We consider 25-30 momenta in the range $0.8 \leq (a p)^2 \leq 7$. The momenta have the following form:
\begin{equation}
(a p) \equiv 2 \pi \left(\frac{n_t + 1/2}{T/a}, \frac{n_x}{L/a}, \frac{n_y}{L/a}, \frac{n_z}{L/a}\right), 
\end{equation}
where $n_i \in \mathbb{Z}$ and $L \ (T)$ denotes the spatial (temporal) extent of the lattice. To minimize rotational $O(4)$ breaking lattice artifacts, we choose momenta close to the body-diagonal direction by imposing $\sum_{\mu} p_\mu^4/(\sum_\mu p_\mu^2)^2 < 0.3$. Additionally, we improve our nonperturbative estimates by subtracting one-loop lattice artifacts from both $Z_q$ and $\Lambda_{\mathcal{O}}$. The artifacts are computed in lattice perturbation theory to all orders in the lattice spacing by extending the improvement program of our group (see Ref.~\cite{Alexandrou:2015sea}) to the two- and three-derivative operators. More details are given in Appendix~\ref{app:artifacts}. The reduction of lattice artifacts and $O(4)$ breaking effects lead to a milder and smoother dependence of the renormalization functions on $a^2 \mu_0^2$ (see Fig.~\ref{fig:Zfactors}).

After chiral extrapolation and artifact subtraction, the resulting renormalization functions are converted to the $\overline{\rm MS}$ scheme and evolved at the reference scale 2 GeV, using an intermediate renormalization group invariant (RGI) scheme~\cite{Gockeler:2010yr} defined in continuum perturbation theory, as follows:
\begin{equation}
Z_{X}^{\overline{\rm MS}} (2\,{\rm GeV}) =  \frac{\Delta Z_{X}^{{\rm RGI}, {\rm RI}'} (\mu_0)}{\Delta Z_{X}^{{\rm RGI}, \overline{\rm MS}}(2\,{\rm  GeV})} Z_{X}^{{\rm RI}'} (\mu_0), 
     \label{eq:Conv}
\end{equation}
for $X={\rm VDD}, {\rm VDDD}$. The quantity $\Delta Z_{X}^{{\rm RGI}, {\mathcal S}}(\mu)$ is expressed in terms of the $\beta$-function and the anomalous dimension of the operators $\mathcal{O}^{\{\mu\nu\rho\}}$ and $\mathcal{O}^{\{\mu\nu\rho\sigma\}}$ in the scheme $\mathcal{S}$, which can be derived to four loops in perturbation theory by using the results of Refs.~\cite{Gracey:2006zr,Baikov:2015tea,Herzog:2018kwj}. In Appendix~\ref{app:conversion_functions}, we provide the explicit perturbative expressions.

Next, we apply a linear fit in $a^2 \mu_0^2$ to eliminate any residual dependence on the RI$'$-MOM scale resulting from discretization effects. Figure~\ref{fig:Zfactors} displays the momentum fits of $Z_{\rm VDD}$ and $Z_{\rm VDDD}$ at a selected fit range. The plots contain data with and without subtracting one-loop artifacts in order to illustrate the benefit of our subtraction method. We found that the subtraction procedure improves the data significantly, leading to a smaller slope and thus to smaller dependence on the initial scale. 

We employ several fit ranges within $2 \leq (a\mu_0)^2 \leq 7$. The extrapolated values at $\mu_0 = 0$ from all fits are combined using model averaging with AIC weights. Momenta with $(a\mu_0)^2 < 2$ are excluded from the analysis, as they may suffer from significant hadronic contamination, and the perturbative conversion is not reliable in this low-momentum region. In Fig.~\ref{fig:AIC}, we show the results from all fits together with the AIC-averaged value. The final estimates for $Z_{\rm VDD}$ and $Z_{\rm VDDD}$ are:
\begin{eqnarray}
    Z_{\rm VDD}^{\overline{\rm MS}} \, (2 \, {\rm GeV}) &=& 1.4639(11)(03), \\
    Z_{\rm VDDD}^{\overline{\rm MS}} \, (2 \, {\rm GeV}) &=& 1.8831(19)(10).
\end{eqnarray}
The number in the first (second) parenthesis corresponds to the statistical (systematic) uncertainty. The systematic uncertainty is determined from the AIC procedure.

\section{Results and discussion}\label{sec:results}
After multiplicatively renormalizing  the bare results, we show in  Table~\ref{tab:moments} the extracted moments renormalized in the $\overline{\mathrm{MS}}$ scheme at  $\mu = 2\,\mathrm{GeV}$. 

\begin{table}[h!]
\centering
\begin{tabular}{lccc}
\hline\hline
 & $\langle x \rangle$ & $\langle x^{2} \rangle$ & $\langle x^{3} \rangle$ \\
\hline
$u^{\pi}$ & 0.2194(22)  & 0.1021(34)  & 0.079(25) \\
$u^{K}$   & 0.2151(12)  & 0.0925(11)  & 0.0557(54) \\
$s^{K}$   & 0.30081(67) & 0.14312(86) & 0.0881(41) \\
\hline\hline
\end{tabular}
\caption{Mellin moments at $\mu = 2\,\mathrm{GeV}$ in $\overline{\mathrm{MS}}$ scheme using only connected contributions. $\langle x \rangle$ is the connected part from Ref.~\cite{ExtendedTwistedMass:2024kjf}.}
\label{tab:moments}
\end{table}

Using the results for the individual moments, we compute various ratios of moments. The ratios in Table~\ref{tab:momentratios}  suggest that the pion moments tend to decrease less with the order of the moment, as compared to the kaon moments. However, the uncertainties of the pion ratios involving $\langle x^3 \rangle$ are large to make a definite conclusion.   Given this suggested tendency of the ratios,  one may  expect the pion PDF to fall more slowly at somewhat  higher $x$-values than the light and strange in the kaon.

\begin{table}[h!]
\centering
\begin{tabular}{lccc}
\hline\hline
& $\biggl.\biggr.u^\pi$ & $u^K$ & $s^K$\\
\hline
$\biggl.\biggr.\frac{\langle x^2 \rangle}{\langle x\rangle}$ & 0.466(16) & 0.430(5) & 0.475(3) \\
$\biggl.\biggr.\frac{\langle x^3 \rangle}{\langle x\rangle}$ & 0.36(11) & 0.259(25) & 0.293(14) \\
$\biggl.\biggr.\frac{\langle x^3 \rangle}{\langle x^2\rangle}$ & 0.77(24) & 0.602(58) & 0.616(30)\\ 
\hline\hline
\end{tabular}
\caption{Ratios of the higher Mellin moments, using the jackknife data leading to the values of Table~\ref{tab:moments}.}
\label{tab:momentratios}
\end{table}

\begin{table}[t]
    \centering
    \begin{tabular}{lccc}
        \hline\hline
        & $\langle x^n \rangle^{\pi}_{u}/\langle x^n \rangle^{K}_{u}$ 
        & $\langle x^n \rangle^{\pi}_{u}/\langle x^n \rangle^{K}_{s}$ 
        & $\langle x^n \rangle^{K}_{u}/\langle x^n \rangle^{K}_{s}$ \\
        \midrule
        $n=1$ & 1.020(12) & 0.7292(76) & 0.7151(46) \\
        $n=2$ & 1.104(34) & 0.714(23) & 0.6466(79) \\
        $n=3$ & 1.41(41) & 0.89(27) & 0.632(67) \\
        \hline\hline
    \end{tabular}
    \caption{Ratios of the Mellin moments between pion and kaon, using the jackknife data leading to the values of Table~\ref{tab:moments}.
    }
    \label{tab:ratiosmoments2}
\end{table}

{In Table~\ref{tab:ratiosmoments2}, the first column compares the ratios of moments of the light quarks in the pion and kaon. The three mean values are larger than unity.  This behavior is consistent with the  observation in Table~\ref{tab:momentratios} suggesting that the PDF of the pion falls more slowly for higher $x$ values than the $u$-quark PDF of the kaon. The second column gives the ratios of the pion $u$-quark to those of the kaon $s$-quark. These ratios point to larger overall moments for the $s$-quark in the kaon, indicating that it carries a larger fraction of its momentum due to  the large mass of the strange quark. Finally, the third column shows the ratios of moments for the $u$-quark and the $s$-quark in the kaon. In the exact SU$(3)$ symmetric limit, these ratios should be equal to 1. We therefore estimate the SU$(3)$ symmetry breaking from the deviation from 1. We see a clear SU$(3)$ symmetry breaking of all moments of about $30-40\%$ for the second and third moments with the breaking being larger for the third moment. Whether this increasing SU(3) breaking trend applies also for the fourth moment is unclear given the present uncertainties.  }

\section{PDF Reconstruction}\label{sec:PDF_reconstruction}

Since we compute only the connected contributions, and assuming all sea quark effects are small, one can use the four lower Mellin moments to estimate the valence PDFs using the moments of Table~\ref{tab:moments}. We employ the commonly used parametrization of PDFs given by
\begin{equation}
    q^{M}_{f}(x) = N x^\alpha (1-x)^\beta (1 + \gamma x),
\label{eq:functional_form}
\end{equation}
where $\alpha$, $\beta$ and $\gamma$ are adjustable parameters and $N$
is a  normalization constant determined from the first moment. Integrating $q_f^M(x)$ using the parametrization of Eq.~(\ref{eq:functional_form}), we obtain
\begin{equation}
    N = \frac{1}{B(\alpha + 1, \beta + 1) + \gamma B(2 + \alpha, \beta + 1)},
\end{equation}
where $B(x,y)$ is the $\beta$ function. 
The $n{\rm th}$ moment of the PDF determined by integrating over $x$  is given by
\begin{equation}
    \langle x^{n-1} \rangle = N \int_0^1 x^{n-1} x^\alpha (1-x)^\beta (1 + \gamma x)dx,
\end{equation}
which yields
\begin{equation}
    \langle x^{n-1} \rangle = \frac{\left( \prod_{i = 1}^{n-1} (\alpha + i) \right) \left( 1 + n +\alpha + \beta + (n + \alpha) \gamma \right)}{\left( \prod_{i = 1}^{n-1} (2 + \alpha + \beta + i) \right)\left( 2 +\alpha + \beta + (1 + \alpha) \gamma \right)}.
    \label{eq:moments_formula}
\end{equation}

Since experimental data analysis suggest that the inclusion of $\gamma$ does not improve the description of the data~\cite{Barry:2025wjx}, we consider two fitting scenarios: the first a two-parameter form with $\gamma$ set to zero, the second a three-parameter form in which $\gamma$ is left as a free parameter. For each quark flavor, the three Mellin moments given in Table~\ref{tab:moments} are fitted using Eq.~\eqref{eq:moments_formula}. Correlations are taken into account using the covariance matrix for the fits. The resulting values of the parameters $\alpha, \beta$ and $\gamma$ are reported in Table~\ref{tab:first_fits}.

\begin{table}[h!]
\centering
\caption{Results for the fit parameters  of the parametrization given in   Eq.~(\ref{eq:functional_form}).}
\begin{tabular}{lcccc}
\hline\hline
Fit type & $\alpha_\pi^u$ & $\beta_\pi^u$ & $\gamma_\pi^u$ & $\chi^2/$ d.o.f. \\
\hline 
2-parameter & -0.528(49) & 0.68(16) & 0 & 0.52 \\
3-parameter & -0.496(48) & 0.14(17) & -0.816(15) & ... \\ %0.49
\hline\hline
Fit type & $\alpha_K^u$ & $\beta_K^u$ & $\gamma_K^u$ & $\chi^2/$ d.o.f. \\
\hline 
2-parameter & -0.434(23) & 1.065(76) & 0 & 0.67 \\
3-parameter & -0.406(23) & 0.528(77) & -0.7853(50) & ... \\ %0.55
\hline\hline
Fit type & $\alpha_K^s$ & $\beta_K^s$ & $\gamma_K^s$ & $\chi^2/$ d.o.f. \\
\hline 
2-parameter & -0.107(21) & 1.076(47) & 0 & 1.31 \\
3-parameter & -0.058(21) & 0.549(48) & -0.8013(28) & ... \\ %1.05
\hline\hline
\end{tabular}
\label{tab:first_fits}
\end{table}

\begin{figure}[h!]
\includegraphics[width=0.5\linewidth]{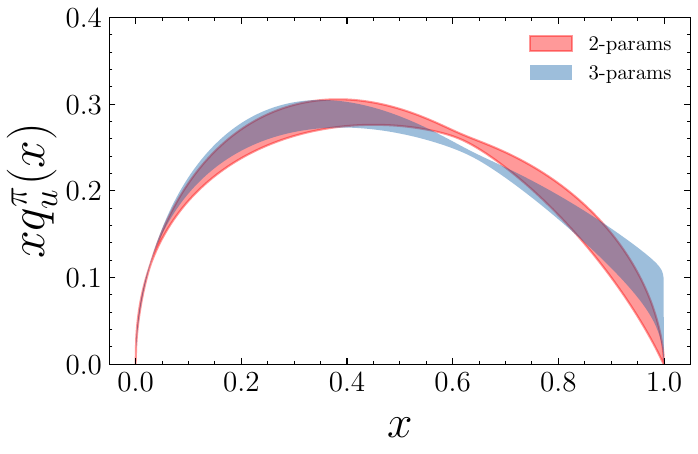}\\
\includegraphics[width=\linewidth]{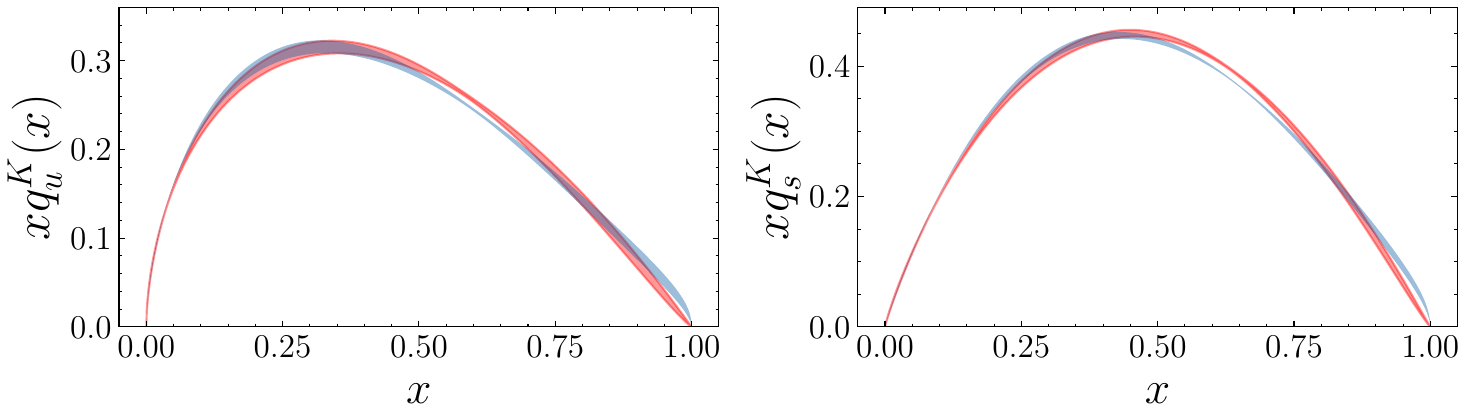}
    \caption{$x$-dependence of the reconstructed PDFs at 2 GeV for $xq^{\pi}_{u}(x)$ (top panel), $xq^{K}_{u}(x)$ (bottom left panel) and $xq^{K}_{s}(x)$ (bottom right panel). Our curves using a two-parameter approach are shown in blue and the three-parameter one in red.}
    \label{fig:2_3_params}
\end{figure}

Using the extracted parameters $\alpha$ and $\beta$ we reconstruct the valence PDFs. The results are shown in Fig.~\ref{fig:2_3_params}. As can be seen, the valence pion PDF is broader than the corresponding kaon PDF consistent with what the higher moment ratios indicated. Allowing for $\gamma\neq0$ we redetermine the parameters the values of which are included in Table~\ref{tab:first_fits}. While the  $\alpha$ values obtained are in agreement, the values of $\beta$ change considerably and the magnitude of $\gamma$ is large and negative. Since both $\beta$ and $\gamma$ parameters affect mostly the large-$x$ dependence, their values are correlated. Comparing the extracted PDFs in Fig.~\ref{fig:2_3_params}, we indeed see that the biggest deviation is seen for $x>0.4$. To further investigate this behavior, we compute the effective $\beta$ parameter, defined as
\begin{equation}
\beta_{\rm eff}\equiv \frac{\partial \text{log}|q(x)|}{\partial\text{log}(1-x)} = \beta -\alpha\frac{1-x}{x}-\gamma\frac{1-x}{1+\gamma x} ,
\label{Eq:beta_eff}
\end{equation}
which describes the effective power of the $1-x$ falloff 
in the large $x$ region. 

\begin{figure}[h!]
    \centering
    \includegraphics[scale = 0.7]{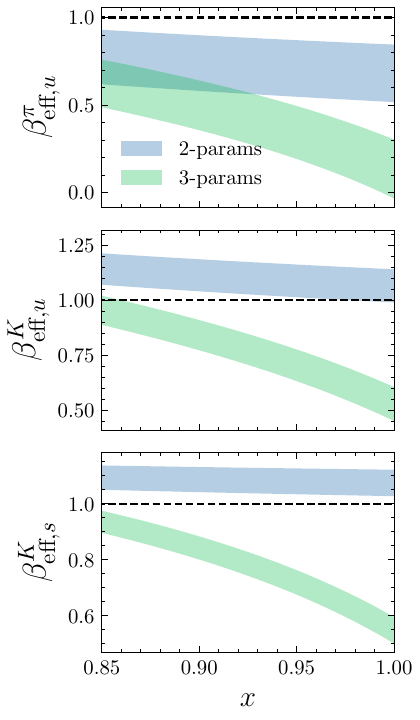}
    \caption{$\beta_{\text{eff}}$ computed using the parameters in Table~\ref{tab:first_fits}. From top to bottom, we show the $\beta_{\text{eff}}$ for the $u$-quark in the pion,  the $u$-quark in the kaon and $s$-quark in the kaon for each reconstructed PDF. The blue band corresponds to the two-parameter reconstruction and the green band to the three-parameter reconstruction. The dashed line at 1 is for reference.}
    \label{fig:eff_beta}
\end{figure}

The results for the $\beta_{\rm eff}$ are shown in Fig.~\ref{fig:eff_beta}. As can be seen, $\beta_{\rm eff}$ shows for the three-parameter fits a strong $x$ dependence in the interval $0.85<x<1$, whereas for the two-parameter fit remains close to unity within errors. This may indicate that with the available moments, the three-parameter form is not sufficiently 
constrained, and a strong $x$ dependence is induced in
$\beta_{\rm eff}$ by the last term in Eq.~\eqref{Eq:beta_eff}.
 In what follows, we will determine the PDFs  using the two-parameter fit
 and take as a systematic error the difference between the mean value from the two- and three-parameter fits. The total error is obtained by adding  the statistical and systematic in quadrature,   $\sigma_{\rm tot} = \sqrt{\sigma_{2 \rm param}^2 + |q_{2 \rm mean} - q_{3 \rm mean}|^2}$, where $\sigma_{2 \rm param}$ is the error of the two-parameter reconstruction and $q_{2 \rm mean}$ and $q_{3 \rm mean}$ the  mean value using the two- and three-parameter reconstructions, respectively.%
\begin{figure}[h!]
    \centering
    \includegraphics[width=0.85\linewidth]{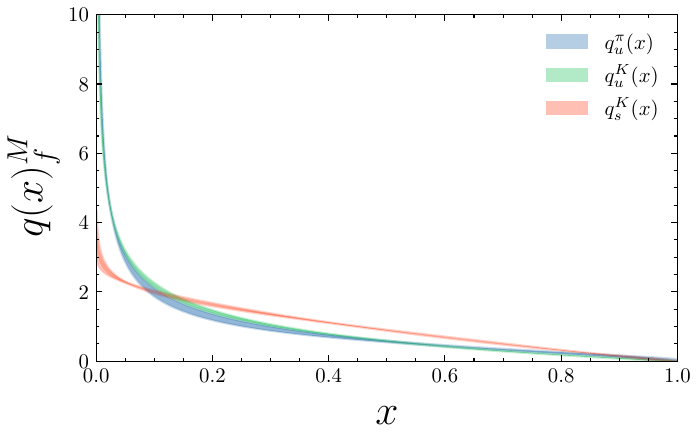}\\
    \includegraphics[width=0.85\linewidth]{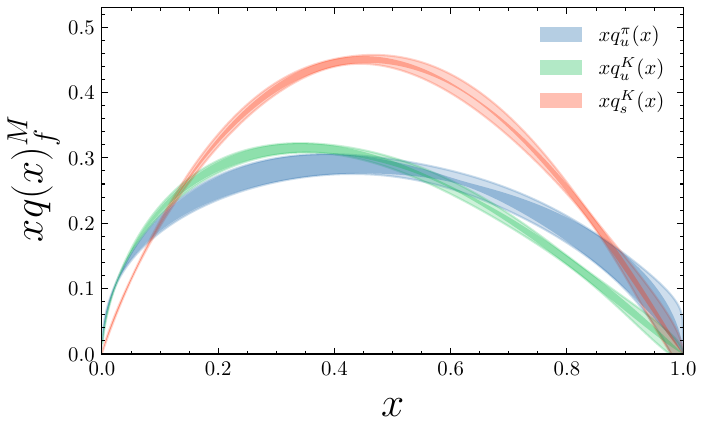}
    \caption{Comparison of the reconstructed PDFs and $q(x)^\pi_u$ (blue band), $q(x)^K_u$ (green band) and $q(x)^K_s$ (red band) at 2 GeV. The upper panel shows the PDFs $q(x)^M_f$ while the lower panel shows the corresponding densities $xq(x)^M_f$. All the bands use the values given in Table~\ref{tab:first_fits} for the two-parameter reconstruction. The outer lighter bands represents the total error from the two- and three-parameter reconstructions, computed by adding in quadrature the uncertainty of the two-parameter result and the difference between mean values of the two- and three-parametrization fits.}
    \label{fig:xqx_all}
\end{figure}

In Fig.~\ref{fig:xqx_all}, we show a comparison of valence pion and kaon PDFs. Comparing the PDFs for the $u$-quark in the pion and kaon, we observe that the pion PDF is broader as predicted by the ratios of moments. This is also reflected in the $\beta_{\rm eff}$ of each PDF, which for the pion remains below unity in the large-$x$ range, whereas for $u$-quark contribution in the kaon it is larger and above unity. For the kaon $s$-quark PDF, the distribution peaks at a larger $x$ value than the $u$-quark PDF in both pion and kaon. This can be quantified by computing the peak of the distributions from
\begin{equation}
    x_{\text{peak}} = \frac{\alpha+1}{\alpha+\beta+1}\, .
\end{equation}
Using  our fitted parameters,
we find $x_{\text{peak},u}^\pi = 0.410(34)$ for the pion PDF, $x_{\text{peak},u}^k= 0.3469(70)$ for kaon $u$-quark PDF, and $x_{\text{peak},s}^K= 0.4535(49)$ for kaon  $s$-quark PDF. These values  confirm that the  $s$-quark distribution has larger support at larger $x$ values as compared with $u$-quark in the kaon. The $s$-quark PDF shows a faster falloff for small values of $x$, consistent with the smaller value of the corresponding $\alpha$ parameter.

\section{Comparison with other results}
\label{sec:comparison_with_other_determinations}

In this section, we compare  our results with other recent lattice-QCD calculations as well as with phenomenology analyses and model calculations. To allow a meaningful comparison, results are quoted at the renormalization scale $\mu=2$~GeV in the $\overline{\rm MS}$ scheme, except for Figs.~\ref{fig:kaon_recons} and \ref{fig:kaon_over_pion} where the moments are evolved to $\mu=5.2$~GeV for the PDF reconstructions.
\begin{figure}[h!]
    \centering
   \includegraphics[width=0.5\textwidth]{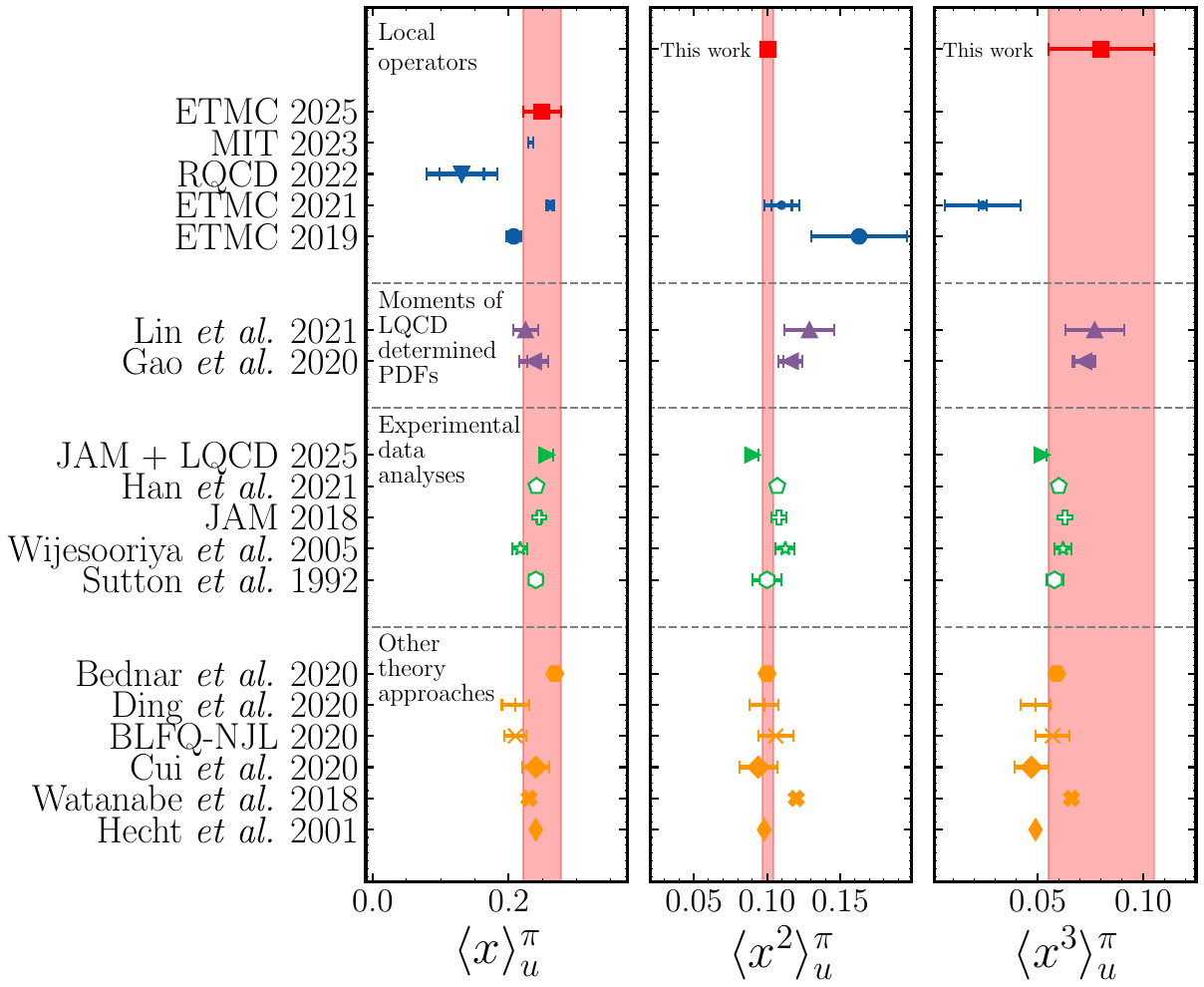}
    \caption{Results on  the pion Mellin moments $\langle x \rangle_u^\pi$ (left panel), $\langle x^2 \rangle_u^\pi$ (middle panel) and  $\langle x^3 \rangle_u^\pi$ (right panel).  We group results  in four categories: i) Lattice QCD using matrix elements of   local operators  (red square and band)~\cite{ExtendedTwistedMass:2024kjf} and from Refs.~\cite{Hackett:2023nkr, Loffler:2021afv,Alexandrou:2020gxs,Alexandrou:2021mmi,Oehm:2018jvm} (blue symbols);  ii) from moments of lattice QCD-determined PDFs~\cite{Lin:2020ssv,Gao:2020ito} (purple symbols); iii) extracted from the analysis of experimental results~\cite{Han:2020vjp,Barry:2018ort,Wijesooriya:2005ir,Sutton:1991ay} (open green symbols) and filled green triangle when lattice QCD input is included~\cite{Barry:2025wjx}; and iv) from other theory approaches~\cite{Bednar:2018mtf,Ding:2019qlr,Lan:2019rba,Watanabe:2017pvl,Cui:2020tdf,Hecht:2000xa} (orange symbols).  The references are given in the order the symbols appear from top to bottom.
    }
    \label{fig:comp_pion}
\end{figure}
\begin{figure}[h!]
    \centering
   \includegraphics[width=\linewidth]{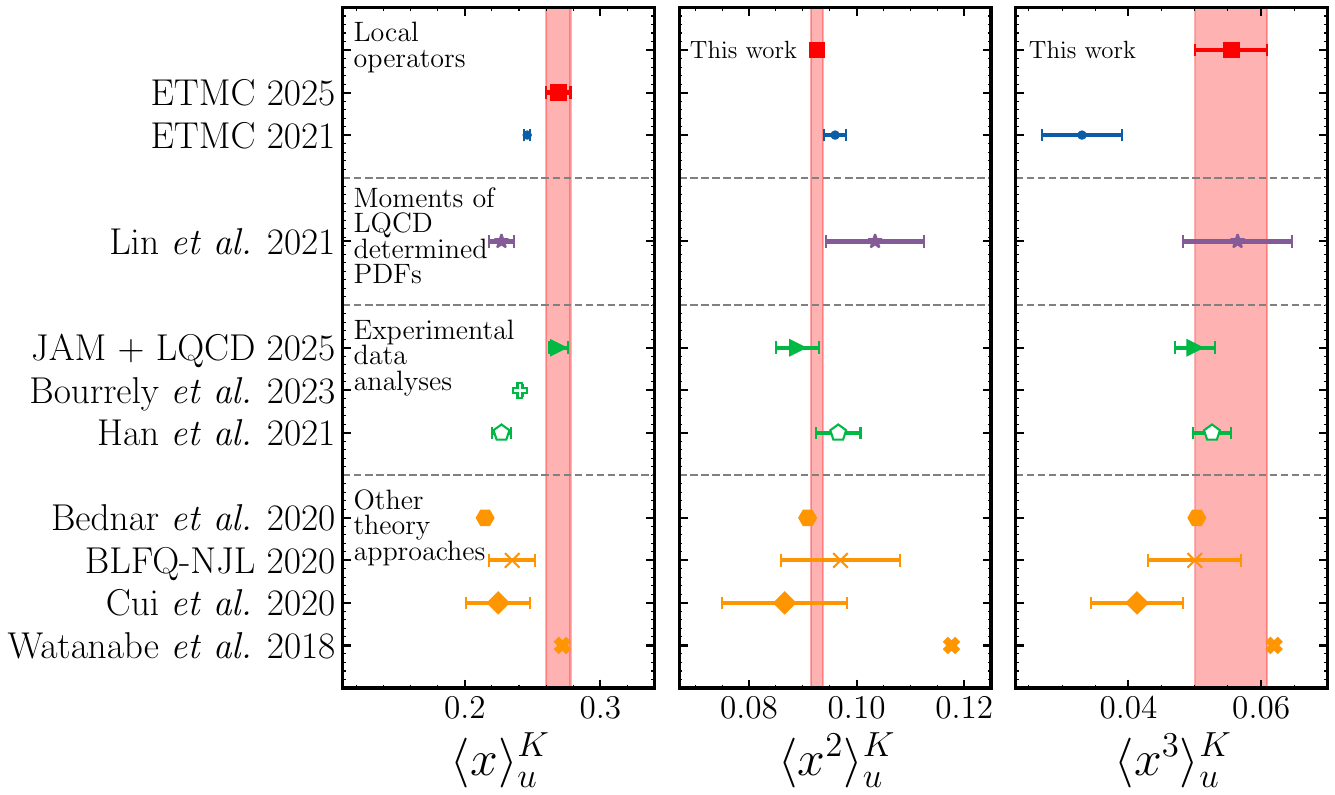}
    \caption{Results on  the kaon $u$-quark Mellin moments $\langle x \rangle_u^K$ (left panel), $\langle x^2 \rangle_u^K$ (middle panel) and  $\langle x^3 \rangle_u^K$ (right panel). We use the same notation as in Fig.~\ref{fig:comp_pion}. The results of the  analysis of experimental data are from Ref.~\cite{Barry:2025wjx, BOURRELY2024138395} from other theory approaches from Refs.~\cite{Bednar:2018mtf,Lan:2019rba,Watanabe:2017pvl,Cui:2020tdf}.}
    \label{fig:comp_kaon_u}
\end{figure}
\begin{figure}[h!]
    \centering
   \includegraphics[width=\linewidth]{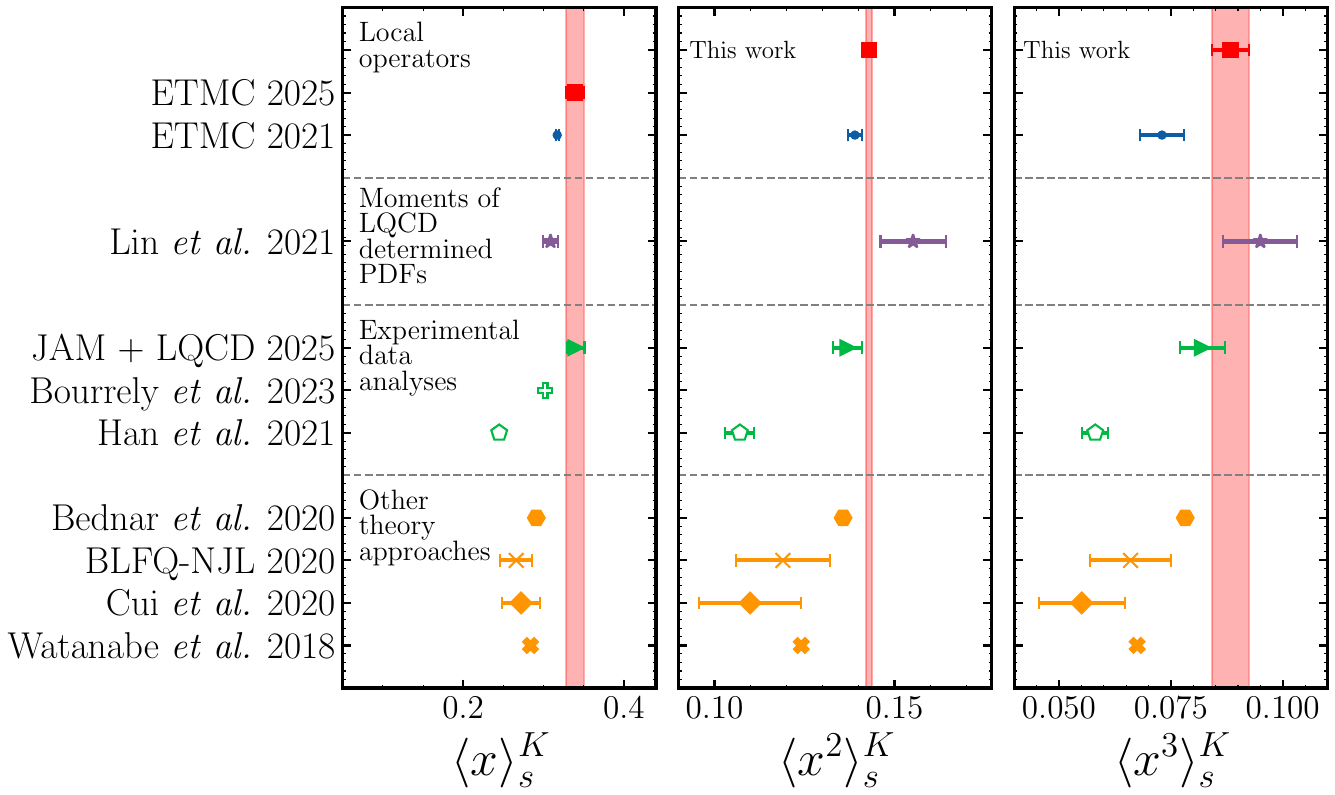}
    \caption{Results on  the kaon $s$-quark Mellin moments $\langle x \rangle_s^K$ (left panel), $\langle x^2 \rangle_s^K$ (middle panel) and  $\langle x^3 \rangle_s^K$ (right panel). We use the same notation as in Fig.~\ref{fig:comp_pion}.}
    \label{fig:comp_kaon_s}
\end{figure}

\subsection{Pion moments}

In Fig.~\ref{fig:comp_pion} we compile results for the second, third, and fourth Mellin moments of the pion. We separated the results in Fig.~\ref{fig:comp_pion} into four groups: i)  lattice QCD results determined from  matrix elements of local operators as in this study,; ii) moments determined in lattice QCD within the quasi/pseudo PDFs and large momentum effective theory (LaMET)~\cite{PhysRevLett.110.262002}; iii)  moments determined from experimental data; and iv) model calculations. 
Note that for $\langle x\rangle_u^\pi$ we quote instead of the $\langle x\rangle$-value from this work the result from Ref.~\cite{ExtendedTwistedMass:2024kjf}, which includes disconnected contributions and is extrapolated to the continuum limit at physical pion mass.

 In general, we observe reasonable agreement for all three pion moments among the various computations, keeping in mind that most lattice QCD computations for the third and fourth moments only have statistical errors. This is the case for example with the earlier ETMC 2021 result, where we found a smaller value by 1 s.d. for $\langle x^3\rangle$  than our present calculation.
In the following, we compile more details on the individual calculations.  Lattice QCD calculations based on local operators there are: i) the MIT group~\cite{Hackett:2023nkr} obtained using an $N_f = 2 + 1$ clover-improved fermion ensemble with  $m_\pi \approx 170$~MeV and lattice spacing $a=0.091$~fm. The value shown includes disconnected contributions; 
ii) the RQCD group using 26 $N_f = 2 + 1$ clover-improved ensembles
with five different lattice spacings ranging from $0.1$~fm to $0.05$~fm, 
and pion masses from $214$~MeV to $420$ MeV. They compute both connected and disconnected contributions, take the continuum limit and extrapolate to the physical pion mass~\cite{Loffler:2021afv};
iii) earlier ETMC work~\cite{Oehm:2018jvm} using 13 $N_f=2+1+1$ twisted mass fermion ensembles with three lattice spacings and heavier than physical pion masses. Their result is obtained after 
chiral and continuum extrapolations. No disconnected contributions were included. 
While all of the above estimate $\langle x\rangle_u^\pi$, only ETMC provides results of $\langle x^2\rangle_u^\pi$ and $\langle x^3\rangle_u^\pi$  using local operators~\cite{Alexandrou:2020gxs,Alexandrou:2021mmi}. 
An alternative lattice approach uses nonlocal operators; There are i) Ref.~\cite{Lin:2020ssv}, where three ensembles generated by the
MILC Collaboration with two different lattice spacings and  
heavier than physical pion mass were used. After performing a chiral and continuum extrapolation of the PDFs the moments are extracted by integrating the fitted PDFs. ii) In Ref.~\cite{Gao:2020ito} a similar analysis was performed using two ensembles of highly improved staggered sea quarks by MILC
and clover-improved valence quarks with hypercubic smearing with lattice spacings $a=0.06$~fm and $0.04$~fm and pion mass of 300~MeV. Besides using the quasi-PDF approach, they also used short-distance factorization or  the pseudo-PDF approach~\cite{PhysRevD.96.094503} to extract the moments based on two ensembles with two different lattice spacings, and a pion mass of 300~MeV. They estimated continuum extrapolated values using their results at two lattice spacings. iii) Recently, results on these moments were obtained using one $N_f=2+1+1$ ensemble of maximally twisted-mass clover fermions with $m_\pi\simeq260$~MeV and $a=0.0934$~fm~\cite{Miller:2026hza}. 
There are several phenomenological analyses of experimental data available shown , which yielded results for all three pion valence moments. The most recent one is the one by the  JAM Collaboration where lattice-QCD input for all three moments was used. In addition, there are model calculations based on the Dyson-Schwinger equations~\cite{Bednar:2018mtf, Ding:2019qlr, Cui:2020tdf, Hecht:2000xa}, light-front quantization~\cite{Lan:2019rba}, and chiral constituent quark models~\cite{Watanabe:2017pvl}. 

\subsection{Kaon moments}

Compared to the pion, the kaon PDFs are considerably less studied. Experimental data on kaon PDFs is more than 40 years old and limited. Recent phenomenological analyses have begun to revisit kaon PDFs~\cite{BOURRELY2024138395,Han:2020vjp} including studies that incorporate lattice QCD input~\cite{Barry:2025wjx}. Several model approaches have provided predictions for the kaon valence distributions and their moments~\cite{Bednar:2018mtf,Lan:2019rba,Watanabe:2017pvl,Cui:2020tdf}, yet the number of available determinations remains smaller than in the pion case.

In Figs.~\ref{fig:comp_kaon_u} and \ref{fig:comp_kaon_s}, we show the comparison of our results for $\langle x^n\rangle_u^K$ and $\langle x^n\rangle_s^K$ in the kaon respectively. In the case of $\langle x\rangle_u^K$, our result is compatible with the JAM analysis, which includes lattice QCD input. Results from other experimental analyses tend to find smaller values as do other theoretical calculations. For $\langle x^2\rangle_u^K$, our result is compatible with JAM and the results of Han \textit{et al.}~\cite{Han:2020vjp} as well with three of the other theoretical approaches.  For $\langle x^3\rangle_u^K$ most of the results are compatible with ours, with the exception of the ETMC 2021, which was carried out using one ensemble at $m_\pi=260$~MeV~\cite{Alexandrou:2021mmi}.

Similarly, for the $s$-quark,  our results for the three moments are in agreement with  the JAM analysis, which includes lattice QCD input. Results from other experimental analyses tend to find smaller values as do other theoretical calculations.  Given the spread of values in both phenomenological and other theoretical studies, lattice QCD determinations provide valuable input on the kaon moments. More experimental data on these moments are expected in the next few years from experiments at CERN (AMBER) and the EIC at BNL.

\begin{figure}
    \centering
    \includegraphics[width=0.9\linewidth]{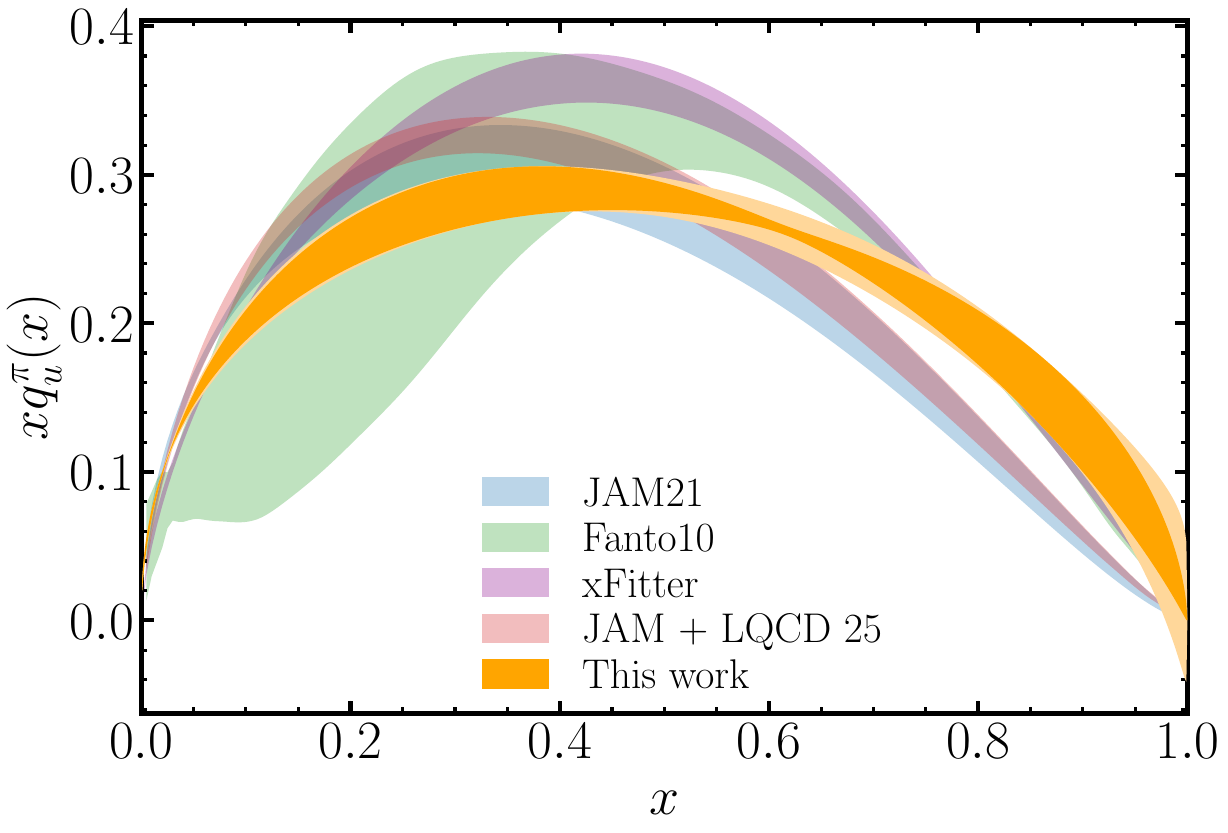}
    \caption{Comparison of the pion PDF $xq^\pi_u(x)$ with the phenomenological extraction at a scale of $2 \,\text{GeV}$ in the $\overline{\text{MS}}$ scheme. The deep orange band corresponds to our result obtained from the two-parameter fits using the moments up to $\langle x^3\rangle$, with the lighter outer band including the systematic uncertainty. The red, blue, green, and magenta bands correspond to the phenomenological determinations by JAM with lattice-QCD input~\cite{Barry:2025wjx} and JAM without lattice-QCD input\cite{Barry:2021osv}, FANTO\cite{Kotz:2025lio}, and xFitter\cite{Novikov:2020snp}, respectively. }
    \label{fig:pion_recons}
\end{figure}

\subsection{Reconstructed PDFs}

In what follows, we  compare 
our reconstructed PDFs with other studies, particularly $xq^M_f(x)$ and the ratio $q(x)^K_u/q(x)^\pi_u$ with the phenomenological ones available in the literature.
Starting with the pion in Figs.~\ref{fig:pion_recons}, the phenomenological results shown are from JAM~\cite{Barry:2021osv,Barry:2025wjx}, FANTO10~\cite{Kotz:2025lio} and xFitter~\cite{Novikov:2020snp}. 
The two JAM determinations overlap with each other over the full $x$-range and show a faster falloff than our result for large-$x$ values. For $x< 0.5$  our results agree with the JAM curves within a standard deviation. The FANTO10 results have significantly larger uncertainties, especially for low-$x$ values, while xFitter seems to have a more
pronounced peak around the $x\approx 0.4$ region. In general our PDF decreases  more slowly for $x>0.5$.
\begin{figure}[h!]
    \centering
    \subfigure[Up quark]{
        \includegraphics[width=0.48\textwidth]{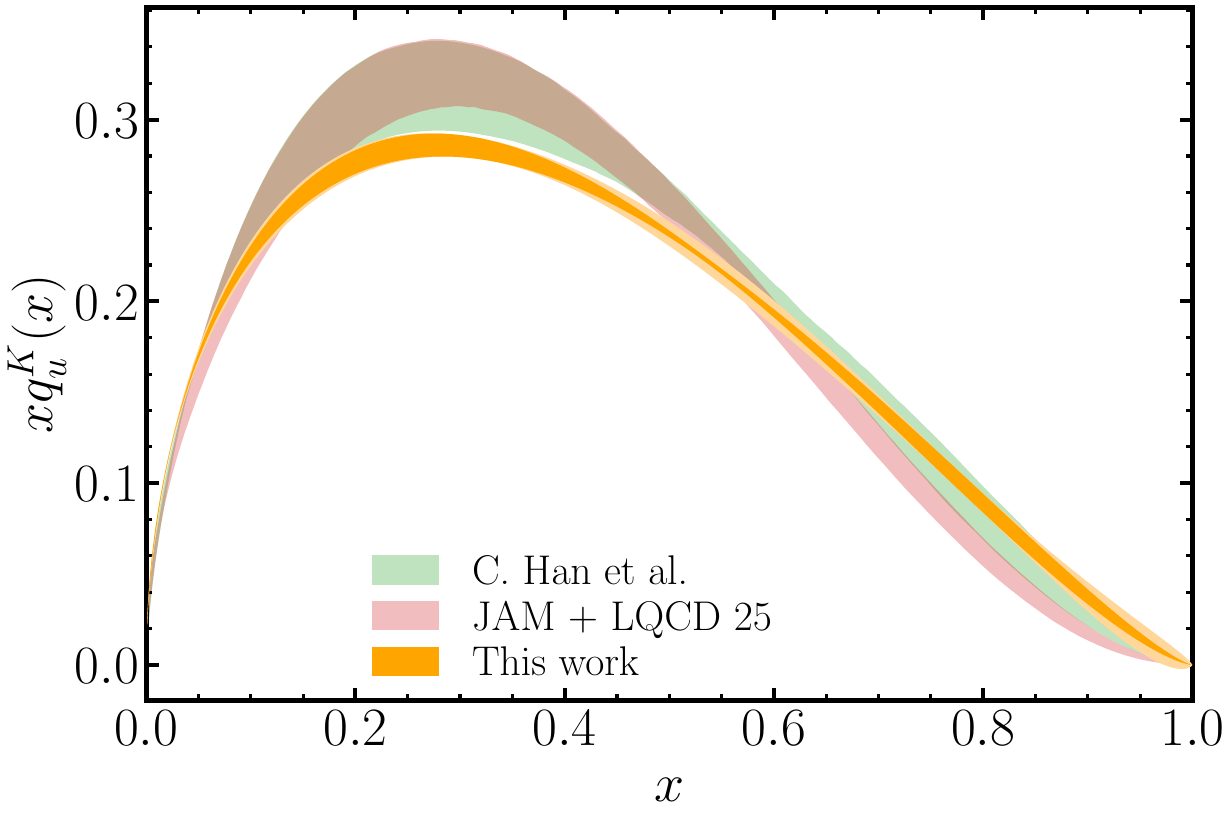}
    }
    \hfill
    \subfigure[Strange quark]{
        \includegraphics[width=0.48\textwidth]{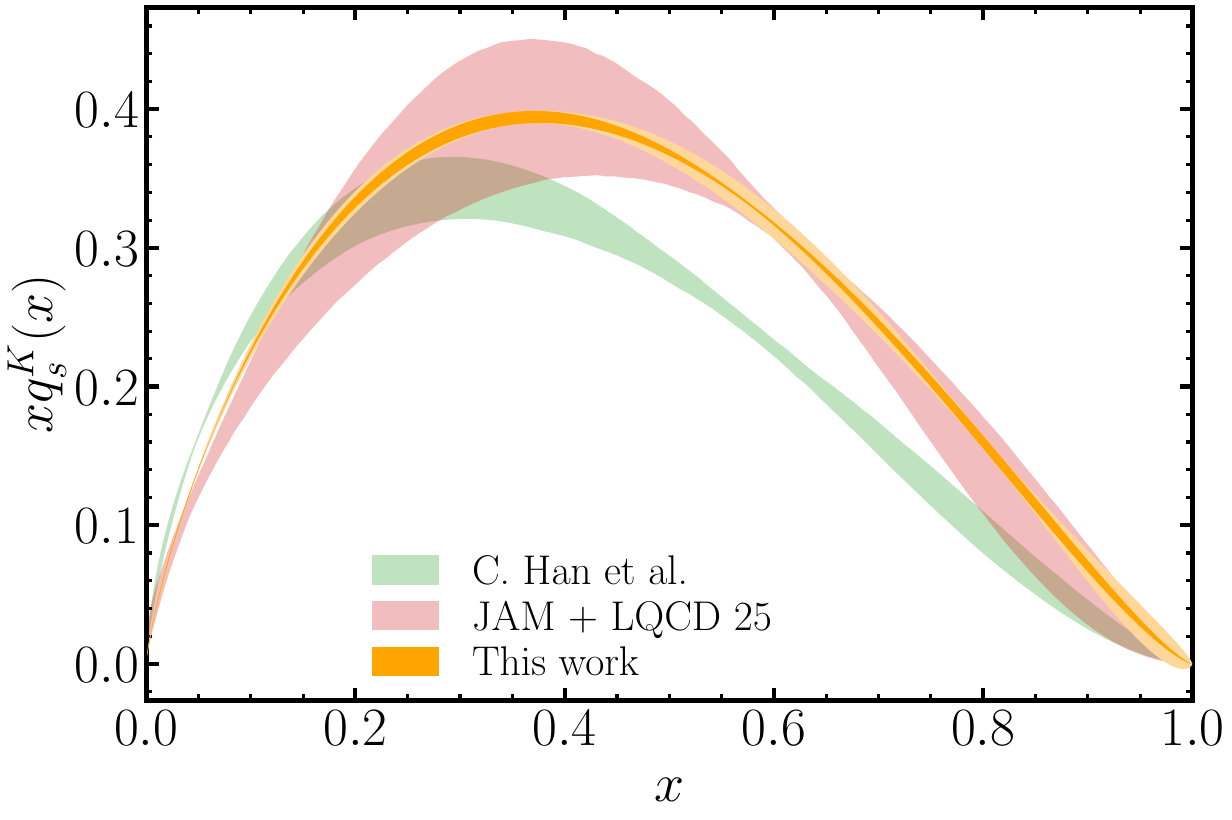}
    }

    \caption{Comparison of our reconstructed PDFs with phenomenological determinations at a scale of $5.2$~GeV in the $\overline{\text{MS}}$ scheme. In the upper panel, we show $xq(x)^K_u$ and in the lower panel  $xq(x)^K_s$. The  deep orange band corresponds to the resulting PDF from the two-parameter fits, using the results in Table~\ref{tab:first_fits} with the lighter outer band including the systematic uncertainty. The green and red bands correspond to the phenomenological extraction from Ref.~\cite{Han:2020vjp}, and by JAM with lattice-QCD input~\cite{Barry:2025wjx}, respectively. }
    \label{fig:kaon_recons}
\end{figure}

In Fig.~\ref{fig:kaon_recons}, we compare our reconstructed kaon PDFs
with JAM~\cite{Barry:2025wjx} and an older analysis from Ref.~\cite{Han:2020vjp},
at the scale of 5.2~GeV. To facilitate this comparison, we first evolve our moments from $2~\mathrm{GeV}$ to $5.2~\mathrm{GeV}$ using the following evolution formula:
\begin{equation}
\langle x^n \rangle^{\overline{\rm MS}} (5.2 , {\rm GeV}) =
\frac{\Delta Z_{{\rm VD}^n}^{{\rm RGI}, \overline{\rm MS}} (2,{\rm GeV})}
{\Delta Z_{{\rm VD}^n}^{{\rm RGI}, \overline{\rm MS}}(5.2,{\rm GeV})}
\langle x^n \rangle^{\overline{\rm MS}} (2 , {\rm GeV}) ,
\end{equation}
and then repeat the PDF reconstruction at the evolved scale. Here, $\Delta Z_{{\rm VD}^n}^{{\rm RGI}, \overline{\rm MS}}(\mu)$ is defined in Appendix~\ref{app:conversion_functions} in terms of the anomalous dimension $\gamma_{{\rm VD}^n}$. For $n=2$ (VDD) and $n=3$ (VDDD), all perturbative ingredients required for the evolution are provided in this Appendix, while for $n=1$ (VD) we refer the reader to Ref.~\cite{Alexandrou:2026lrh}. JAM relies on ETMC input~\cite{ExtendedTwistedMass:2024kjf} 
to constrain the kaon PDFs, which would otherwise be
impossible to determine from experimental data only. There is an overall agreement within errors between our PDF for $xq^K_u(x)$ and the phenomenological results. 
In the case of $xq(x)^K_s$, there is complete agreement between 
our and JAM's results, while the older extraction from Ref.~\cite{Han:2020vjp} shows a visibly different shape. The errors in the JAM PDFs are larger than
ours because besides using the earlier result of Ref.~\cite{ExtendedTwistedMass:2024kjf}, they have inflated the errors by a factor of 2 to capture possible unaccounted systematic uncertainties.
\begin{figure}[h!]
    \centering
    \includegraphics[width=0.9\linewidth]{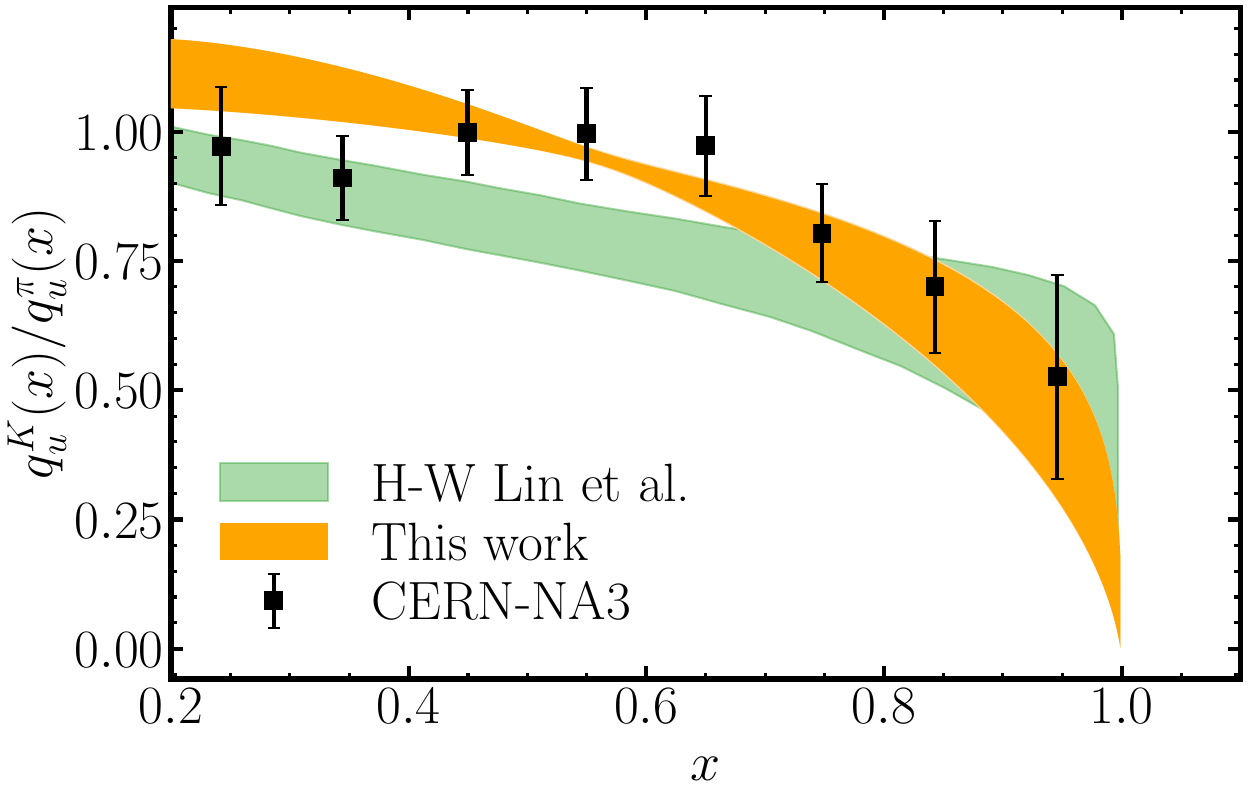}
    \caption{Ratio $q_u^K(x)/q_u^\pi(x)$ as a function of $x$ compared with the lattice result from Lin \textit{et al}. \cite{Lin:2020ssv} and the CERN-NA3 experimental data \cite{Saclay-CERN-CollegedeFrance-EcolePoly-Orsay:1980fhh} at a scale of $5.2$~GeV in the $\overline{\text{MS}}$ scheme. The systematic error is shown but it is too small to be visible. }
    \label{fig:kaon_over_pion}
\end{figure}
In Fig.~\ref{fig:kaon_over_pion}, we show the ratio of PDFs,
$q(x)^K_u/q(x)^\pi_u$ 
along with  results from Ref.~\cite{Lin:2020ssv} computed within lattice QCD using the quasidistribution approach. 
We also show available experimental data from the NA3 experiment at CERN~\cite{Saclay-CERN-CollegedeFrance-EcolePoly-Orsay:1980fhh}. Our results show a good agreement with the experimental data over the
whole $x$ region, while showing some tension with those of Ref~\cite{Lin:2020ssv} in the moderate
to low $x$ region.

\section{Conclusions and Outlook}\label{sec:conclusions}

In this work, we present the computation of 
the Mellin moments $\langle x^2 \rangle$ and $\langle x^3 \rangle$ for the pion and the kaon. We employ one ensemble of
$N_f = 2 + 1 + 1$ twisted mass fermions, with their masses
tuned to reproduce the physical pion mass. The calculation was performed by using local operators choosing indices for our operators  to avoid mixing for both moments. We only include connected contributions for all moments.
Renormalization is carried out nonperturbatively within 
the RI$^\prime$-MOM scheme. Our results for the moments are 
summarized in Table~\ref{tab:moments} at a scale of $2$~GeV 
in the $\overline{\text{MS}}$ scheme. 
The ratios between these three moments indicate SU$(3)$ 
breaking on the level of $30\%-40\%$ with the mean values suggesting a possible increasing trend of breaking for higher moments. Reducing  current uncertainties will allow for this trend to be established conclusively.

The pion and the kaon valence PDFs are reconstructed using the first four
Mellin moments. 
Such reconstructions are performed using a parametric form with two- and three-parameters for the valence PDFs. 
The overall shape of the reconstructed PDF are visually consistent with
each other even if the numerical values of some of the parameters are notably different. The $u$-quark PDF of the pion and kaon show similarities, while the kaon $s$-quark PDF has a more pronounced peak and falls faster both for low and large $x$ values.

Although the reconstructed PDFs with two and three parameters are visually
compatible, their falloff in the large-$x$ region are different. In particular,
we study in Fig.~\ref{fig:eff_beta} the effective power $\beta_\mathrm{eff}$ that 
dictates how the PDFs approach zero as $x\rightarrow 1$.
From our fitted parameters, we conclude that the effective $\beta$-parameter is around $10\% - 30\%$ below unity for the pion PDFs, while 
slightly above unity for the kaon $u$- and $s$-quark PDFs. For comparison, JAM computes $\beta_{\rm eff}$ for each distribution, obtaining $\beta^\pi_{{\rm eff}, u} = 1.16(4)$, $\beta^K_{{\rm eff}, u} = 1.6(2)$ and $\beta^K_{{\rm eff}, s} = 1.2(4)$ for the range $x = 0.7-0.95$.

In future studies, various improvements can be made, which include  the calculation of the higher moments for more ensembles so the continuum limit can be taken. Another interesting expansion of our work is to compute disconnected contributions for the higher moments, which are expected to be smaller, but would allow us to have a better understanding of the complete quark valence and sea contributions, allowing also a reconstruction for such PDFs. Novel approaches may also allow extraction of even higher Mellin moments leading to a better determination of PDFs such a using gradient flow~\cite{Shindler:2023xpd, Francis:2025pgf,z3wr-zk8n} or within the heavy-quark operator product expansion~\cite{PhysRevD.73.014501,Detmold:2022dmw,xs4z-nblq}. In parallel improvements in the direct determination of PDFs, using nonlocal operators allow us to compute higher Mellin moments.  

\section*{Acknowledgments}
This project is partly funded by the European Union’s Horizon 2020 Research and Innovation Programme ENGAGE under the Marie Sklodowska-Curie COFUND scheme with Grant Agreement No. 101034267. C.A., S.B., and G.S. acknowledge partial support from the projects 3D-nucleon (EXCELLENCE/0421/0043) and IMAGE-N (EXCELLENCE/0524/0459), MuonHVP (EXCELLENCE/0524/0017), StrongILA (EXCELLENCE/0524/0001), and partonWF (VISION ERC/0525/0010) funded by the European Regional Development Fund and the Republic of Cyprus through the Cyprus Research and Innovation Foundation. C.A. also acknowledges partial support from the European Joint Doctorate project AQTIVATE funded by the European Commission under the Marie Sklodowska-Curie Doctoral Networks action and Grant Agreement No 101072344. U.W.~acknowledges funding from the Swiss National Science Foundation (SNSF) Project No.~200020\_208222. This project was funded in part by the Deutsche Forschungsgemeinschaft (DFG, German Research Foundation) as a project in the CRC 1639 NuMeriQS -- Project No. 511713970 and under Germany’s Excellence Strategy–EXC 3107–Project-ID 533766364 in the cluster of excellence Color meets Flavor. The authors gratefully acknowledge the Gauss Centre for Supercomputing e.V. (www.gauss-centre.eu) for funding this project by providing computing time through the John von Neumann Institute for Computing (NIC) on the GCS Supercomputers JUWELS, JUWELS Booster~\cite{JUWELS-BOOSTER} and JUPITER Booster at J\"ulich Supercomputing Centre (JSC). We also acknowledge computing
time granted on Piz Daint at Centro Svizzero di Calcolo Scientifico (CSCS) via  Projects No. s849, No. s982, No. s1045, No. s1133 and No. s1197, and access to the LUMI supercomputer through the Chronos program under project IDs CH17-CSCS-CYP and CH21-CSCS-UNIBE.

\section*{Data Availability}
There are no publicly available research data or software supporting this manuscript. Requests for further information or data should be sent to the authors.
\bibliography{biblio}

\appendix

\section{One-loop lattice artifacts in the renormalization factors}
\label{app:artifacts}

The nonperturbative estimates of the renormalization factors $Z_{\rm VDD}$ and $Z_{\rm VDDD}$ are improved by subtracting one-loop lattice artifacts stemming from the renormalization factor of the quark field $Z_q (a p)$ and the ``projected'' vertex functions $\mathcal{V}_{\rm VDD} (a p)$ and $\mathcal{V}_{\rm VDDD} (a p)$, defined as [cf. Eqs.~\ref{ZVDD} - \ref{ZVDDD}]:
\begin{align}
  \mathcal{V}_{\rm VDD} (a p) &= \frac{1}{12} \sum_{\mu<\nu<\rho} {\rm Tr} \left[ \Lambda_{\mathcal{O}^{\{\mu\nu\rho\}}} P^{\mu\nu\rho} \right], \\
  \mathcal{V}_{\rm VDDD} (a p) &= \frac{1}{12} \sum_{\mu<\nu<\rho<\sigma} {\rm Tr} \left[ \Lambda_{\mathcal{O}^{\{\mu\nu\rho\sigma\}}} P^{\mu\nu\rho\sigma} \right].
\end{align}
We summarize below the calculation procedure.

\begin{figure}[ht!]
\centering
\includegraphics[width=\columnwidth]{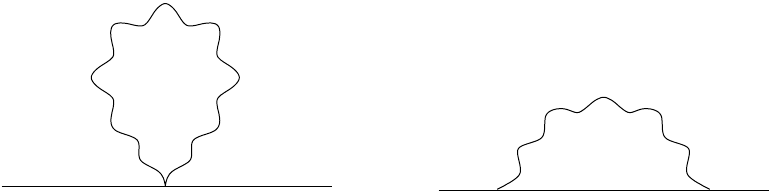} \\
\vspace{1cm}
\includegraphics[width=0.9\columnwidth]{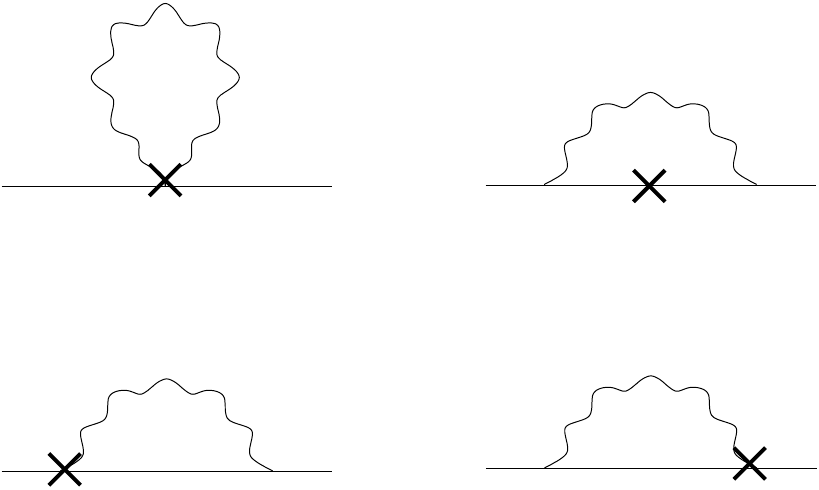}
    \caption{One-loop diagrams contributing to $Z_q$ (top two) and $\mathcal{V}_{\rm VDD}$, $\mathcal{V}_{\rm VDDD}$ (bottom four). Wavy (solid) lines represent gluons (fermions). A cross denotes an insertion of the operator $\mathcal{O}^{\{\mu\nu\rho\}}$ or $\mathcal{O}^{\{\mu\nu\rho\sigma\}}$.}
    \label{fig:Feynman_diagrams}
\end{figure}
We first compute $Z_q$, $\mathcal{V}_{\rm VDD}$, and $\mathcal{V}_{\rm VDDD}$ to one loop in the Landau gauge by considering the Feynman diagrams of Fig.~\ref{fig:Feynman_diagrams}. The calculation is split into two parts. The first part considers the one-loop computation of the three functions at fixed values of $(a p)$, denoted by $Z_q^{(1)} (a p)$, $\mathcal{V}_{\rm VDD}^{(1)} (a p)$, and $\mathcal{V}_{\rm VDDD}^{(1)} (a p)$. These cannot be expressed in a closed analytic form but can be only determined numerically for each value of $(a p)$ employed in our nonperturbative calculation. In the second part, the one-loop functions are expanded over $(a p)$, as follows:\footnote{Since the projector used for the three-derivative operator does not annihilate the second structure in Eq.~\eqref{LVDDD}, $\nu_{{\rm VDDD},0}^{(1)}$ receives an additional contribution proportional to $16/(\sum_\mu p^2/p_\mu^2)$. This contribution approaches unity for momenta close to the body-diagonal direction.}
\begin{align}
    Z_q^{(1)} (a p) =& \frac{g_0^2}{16 \pi^2} [z_{q,0}^{(1)} + \gamma_{q,0} \ln(a p) + \nonumber \\
    & \ \, \mathcal{O} (a^2 p^2, a^2 \frac{\sum_\mu p_\mu^4}{p^2}, a^2 \frac{\sum_\mu p_\mu^6}{(p^2)^2}, \cdots)], \\
   \mathcal{V}_{\rm VDD}^{(1)} (a p) =& \frac{g_0^2}{16 \pi^2} [\nu_{{\rm VDD},0}^{(1)} + \gamma_{{\rm VDD},0} \ln(a p) + \nonumber \\
    & \ \, \mathcal{O} (a^2 p^2, a^2 \frac{\sum_\mu p_\mu^4}{p^2}, a^2 \frac{\sum_\mu p_\mu^6}{(p^2)^2}, \cdots)], \\
    \mathcal{V}_{\rm VDDD}^{(1)} (a p) =& \frac{g_0^2}{16 \pi^2} [\nu_{{\rm VDDD},0}^{(1)} + \gamma_{{\rm VDDD},0} \ln(a p) + \nonumber \\
    & \ \, \mathcal{O} (a^2 p^2, a^2 \frac{\sum_\mu p_\mu^4}{p^2}, a^2 \frac{\sum_\mu p_\mu^6}{(p^2)^2}, \cdots)],
\end{align}
where $\gamma_{q,0} = 0$, and $\gamma_{{\rm VDD},0}$, $\gamma_{{\rm VDDD},0}$ are given in Eqs.~\eqref{app:g_VDD0} and \eqref{app:g_VDDD0}, respectively. We calculate the leading-order terms $\mathcal{O}(a^0 p^0)$ and $\mathcal{O}(\ln(a p))$. The difference between the results from the two parts gives the one-loop artifacts for each function:
\begin{align}
    \Delta Z_q^{(1)} (a p) \equiv&  \, Z_q^{(1)} (a p) -  \frac{g_0^2}{16 \pi^2} [z_{q,0}^{(1)} + \gamma_{q,0} \ln(a p)], \\
    \Delta \mathcal{V}_{\rm VDD}^{(1)} (a p) \equiv& \, \mathcal{V}_{\rm VDD}^{(1)} (a p) - \nonumber \\
    & \, \frac{g_0^2}{16 \pi^2} [\nu_{{\rm VDD},0}^{(1)} + \gamma_{{\rm VDD},0} \ln(a p)], \\
    \Delta \mathcal{V}_{\rm VDDD}^{(1)} (a p) \equiv& \, \mathcal{V}_{\rm VDDD}^{(1)} (a p) - \nonumber \\
    & \, \frac{g_0^2}{16 \pi^2} [\nu_{{\rm VDDD},0}^{(1)} + \gamma_{{\rm VDDD},0} \ln(a p)].
\end{align}

\begin{figure}[ht!]
\centering
\includegraphics[width=\columnwidth]{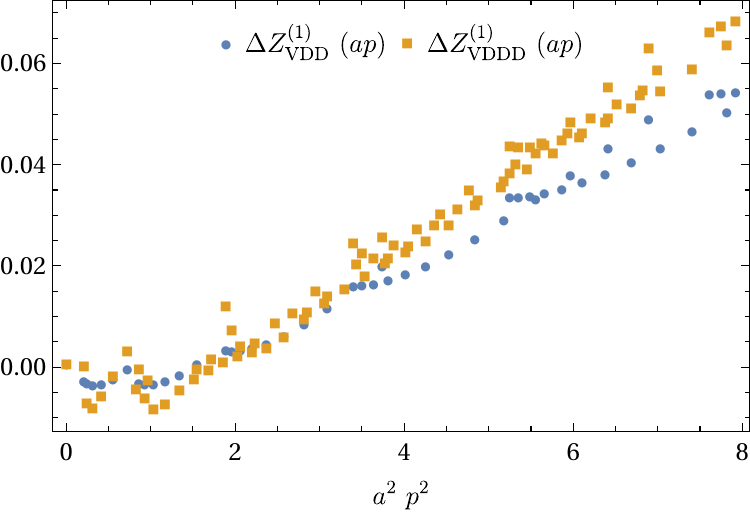} 
    \caption{One-loop lattice artifacts $\Delta Z_{\rm VDD}^{(1)}$ and $\Delta Z_{\rm VDDD}^{(1)}$ as a function of $a^2 p^2$.}
    \label{fig:artifacts}
\end{figure}

The subtraction of artifacts can be done in many ways, e.g., at the level of the projected vertex functions, or at the level of the renormalization factors. Different subtraction methods differ by $\mathcal{O} (g_0^4)$ terms. Below, we provide our selected subtraction method, applied at the level of the renormalization factors, which results in milder dependence on $(a^2 p^2)$, as follows:
\begin{align}
    Z_{\rm VDD}^{\rm impr.} =& Z_{\rm VDD} - \Delta Z_{\rm VDD}^{(1)}, \\
    Z_{\rm VDDD}^{\rm impr.} =& Z_{\rm VDDD} - \Delta Z_{\rm VDDD}^{(1)},
\end{align}
where $\Delta Z_{\rm VDD}^{(1)}$ and $\Delta Z_{\rm VDDD}^{(1)}$ are determined by expanding the ratio $Z_{X}=Z_q/\mathcal{V}_{X}$, with $X=$ VDD or VDDD, to ${\cal O}(g_0^2)$, as follows:
\begin{equation}
Z_{X} = \frac{1+Z_q^{(1)}+{\cal O}(g_0^4)}{1+\mathcal{V}_{X}^{(1)}+{\cal O}(g_0^4)} = 1+\big(Z_q^{(1)}-\mathcal{V}_{X}^{(1)}\big)+{\cal O}(g_0^4)\,,
\end{equation}
so that $Z_{X}^{(1)}=Z_q^{(1)}-\mathcal{V}_{X}^{(1)}$ and
\begin{equation}
\Delta Z_{X}^{(1)}(ap) = \Delta Z_q^{(1)}(ap)-\Delta \mathcal{V}_{X}^{(1)}(ap)\,.
\end{equation}
Figure~\ref{fig:artifacts} shows numerical results for $\Delta Z_{\rm VDD}^{(1)}$ and $\Delta Z_{\rm VDDD}^{(1)}$ as a function of $a^2 p^2$. We set $g_0^2 = 6/\beta$ with $\beta = 1.778$ corresponding to the ensemble used in our calculation. Also, the clover coefficient is set to its tree-level value, $c_{\rm SW} = 1$.

\section{Perturbative conversion functions}
\label{app:conversion_functions}

We provide the four-loop perturbative expressions for the conversion functions $\Delta Z_{X}^{{\rm RGI}, {\mathcal S}}(\mu)$ appearing in Eq.~\eqref{eq:Conv} for the operators $\mathcal{O}^{\{\mu\nu\rho\}}$ ($X={\rm VDD}$) and $\mathcal{O}^{\{\mu\nu\rho\sigma\}}$ ($X={\rm VDDD}$) in both RI$'$-MOM ($S={\rm RI}'$) and $\overline{\rm MS}$ ($S=\overline{\rm MS}$) schemes. They are written in terms of the $\beta$-function and the anomalous dimensions $\gamma_{X}^S$.

Using standard notation, we define (in terms of the renormalization scale $\mu$) the Callan-Symanzik equations satisfied by the renormalized gauge coupling $g^S$ and the renormalized operators in an arbitrary renormalization scheme, denoted by $S$, as follows:
\begin{eqnarray}
    \mu \frac{d}{d\mu} g^S (\mu) &=& \beta^S (g^S(\mu)), \label{beta}\\
    \mu \frac{d}{d\mu} (\mathcal{O}^{\{\mu\nu\rho\}})^S &=& -\gamma_{\rm VDD}^{S} (g^S(\mu)) \ (\mathcal{O}^{\{\mu\nu\rho\}})^S, \\
    \mu \frac{d}{d\mu} (\mathcal{O}^{\{\mu\nu\rho\sigma\}})^S &=& -\gamma_{\rm VDDD}^{S} (g^S(\mu)) \ (\mathcal{O}^{\{\mu\nu\rho\sigma\}})^S. \label{gamma_VDDD}
\end{eqnarray}
The perturbative expansions of the $\beta$ function and the anomalous dimensions $\gamma_{X}$ are defined below in terms of $a (\mu)\equiv (g (\mu))^2/(16 \pi^2)$, as follows:
\begin{eqnarray}
    \beta^S (g) &=& -g \, a \, [\beta_0 + \beta_1 a + \beta^S_2 a^2 + \beta^S_3 a^3 + \mathcal{O} (a^4)], \quad \\
    \gamma^S_{X} (g) &=& a \, [\gamma_{X,0} + \gamma_{X,1}^S \, a + \gamma_{X,2}^S \, a^2 + \gamma_{X,3}^S \, a^3   + \mathcal{O} (a^4)]. \qquad \label{pertexpansion}
\end{eqnarray}
The coefficients $\beta^S_k$ and $\gamma_{X,k}^S$ are extracted in the RI$'$-MOM and $\overline{\rm MS}$ schemes by combining results from Refs.~\cite{Gracey:2006zr,Baikov:2015tea,Herzog:2018kwj}. The following scheme-conversion formulas derived from Eqs.\ref{beta}--\ref{gamma_VDDD} are used:
\begin{eqnarray}
    \beta^{{\rm RI}'} (g^{{\rm RI}'}(\mu)) &=& \beta^{\overline{\rm MS}} (g^{\overline{\rm MS}}(\mu)) \frac{\partial g^{{\rm RI}'}(\mu)}{\partial g^{\overline{\rm MS}}(\mu)}, \\
    \gamma^{{\rm RI}'}_{X} (g^{{\rm RI}'}(\mu)) &=& \gamma^{\overline{\rm MS}}_{X} (g^{\overline{\rm MS}}(\mu)) \nonumber \\
    && \hspace{-1.5cm} + \beta^{\overline{\rm MS}} (g^{\overline{\rm MS}}(\mu)) \,\frac{\partial \ln{C_{X}^{\overline{\rm MS},{\rm RI}'} (\mu)}}{\partial g^{\overline{\rm MS}}(\mu)}, \label{scheme-conversion}
\end{eqnarray}
where $C_{X}^{\overline{\rm MS},{\rm RI}'}= Z_{X}^{\overline{\rm MS}}/Z_{X}^{{\rm RI}'}$. For the specific RI$'$ conditions defined in Eqs. \ref{ZVDD} - \ref{ZVDDD}, $C_{X}^{\overline{\rm MS},{\rm RI}'}$ takes the following form,
\begin{align}
  C_{\rm VDD}^{\overline{\rm MS},{\rm RI}'}  &= C_{q}^{\overline{\rm MS},{\rm RI}'} (\Sigma_1^{\rm VDD})^{\overline{\rm MS}}, \\
  C_{\rm VDDD}^{\overline{\rm MS},{\rm RI}'}  &= C_{q}^{\overline{\rm MS},{\rm RI}'} [(\Sigma_1^{\rm VDDD})^{\overline{\rm MS}} + \frac{16}{\sum_\mu p^2/p_\mu^2} (\Sigma_2^{\rm VDDD})^{\overline{\rm MS}}], 
\end{align}
where $C_{q}^{\overline{\rm MS},{\rm RI}'}= Z_{q}^{\overline{\rm MS}}/Z_{q}^{{\rm RI}'}$ is calculated in Ref.~\cite{Gracey:2003yr}, and $(\Sigma_1^{\rm VDD})^{\overline{\rm MS}}$, $(\Sigma_1^{\rm VDDD})^{\overline{\rm MS}}$, $(\Sigma_2^{\rm VDDD})^{\overline{\rm MS}}$ are extracted from Ref.~\cite{Gracey:2006zr}, to three loops. We provide below the coefficients $\beta^S_k$ and $\gamma_{X,k}^S$ in the Landau gauge and in SU(3). 
\begin{widetext}
\begin{align}
\beta_0 &= 11 - \frac{2}{3}N_f, \\
\beta_1 &= 102 - \frac{38}{3}N_f, \\
\beta_2^{\overline{\rm MS}} &= \beta_2^{{\rm RI}'} = \frac{2857}{2} - \frac{5033}{18}N_f + \frac{325}{54}N_f^2, \\
\beta_3^{\overline{\rm MS}} &= \beta_3^{{\rm RI}'} = \frac{149753}{6} + 3564\,\zeta(3) - \left(\frac{1078361}{162} + \frac{6508}{27}\zeta(3)\right)N_f + \left(\frac{50065}{162} + \frac{6472}{81}\zeta(3)\right)N_f^2
+ \frac{1093}{729}N_f^3, \\
\gamma_{{\rm VDD},0}
&= \frac{100}{9}, \label{app:g_VDD0} \\
\gamma_{{\rm VDD},1}^{\overline{\rm MS}}
&= \frac{34450}{243} - \frac{830}{81}N_f, \\
\gamma_{{\rm VDD},2}^{\overline{\rm MS}}
&= \frac{64486199}{26244}
+ \frac{2200}{81}\zeta(3)  - \left(\frac{967495}{4374}
+ \frac{4000}{27}\zeta(3)\right)N_f
- \frac{2569}{729}N_f^2, \\
\gamma_{{\rm VDD},3}^{\overline{\rm MS}}
&= \frac{69231923065}{1417176}
+ \frac{73641835}{6561}\zeta(3) - \frac{12100}{27}\zeta(4)
- \frac{3669100}{243}\zeta(5) \nonumber\\
&\quad
+ \Big(
-\frac{1978909951}{236196}
- \frac{19276720}{2187}\zeta(3)  + \frac{200200}{81}\zeta(4)
+ \frac{46000}{9}\zeta(5)
\Big)N_f \nonumber\\
&\quad
+ \left(
\frac{3466612}{19683}
+ \frac{24400}{81}\zeta(3)
- \frac{4000}{27}\zeta(4)
\right)N_f^2  + \left(
-\frac{23587}{13122}
+ \frac{1600}{243}\zeta(3)
\right)N_f^3,
\end{align}
\begin{align}
\gamma_{{\rm VDD},1}^{\mathrm{RI}'} &=
\frac{76822}{243} - \frac{562}{27}N_f,
\\
\gamma_{{\rm VDD},2}^{\mathrm{RI}'} &=
\frac{1889349409}{131220}
- \frac{744568}{405}\zeta(3)
+ \left(
-\frac{20589053}{10935}
- \frac{4736}{135}\zeta(3)
\right)N_f
+ \frac{34330}{729}N_f^2,
\\
\gamma_{{\rm VDD},3}^{\mathrm{RI}'} &=
\frac{712102401499}{885735}
- \frac{6011913226}{32805}\zeta(3)
+ \frac{7058540}{243}\zeta(5)
\nonumber \\
&\quad
+ \left(
-\frac{371002768589}{2361960}
+ \frac{25155454}{3645}\zeta(3)
+ \frac{21920}{9}\zeta(5)
\right)N_f
\nonumber \\
&\quad
+ \left(
\frac{828425857}{98415}
+ \frac{96568}{405}\zeta(3)
\right)N_f^2
- \frac{795085}{6561}N_f^3, \\
\gamma_{{\rm VDDD},0}
&= \frac{628}{45}, \label{app:g_VDDD0}
\\
\gamma_{{\rm VDDD},1}^{\overline{\rm MS}}
&= \frac{5241914}{30375}
- \frac{26542}{2025}N_f,
\\
\gamma_{{\rm VDDD},2}^{\overline{\rm MS}}
&= \frac{245787905651}{82012500}
+ \frac{11512}{405}\zeta(3)
- \left(
\frac{726591271}{2733750}
+ \frac{5024}{27}\zeta(3)
\right)N_f
- \frac{384277}{91125}N_f^2,
\\
\gamma_{{\rm VDDD},3}^{\overline{\rm MS}}
&= \frac{1267599127484293}{22143375000}
+ \frac{117362582038}{4100625}\zeta(3)
- \frac{63316}{135}\zeta(4)
- \frac{64357588}{3645}\zeta(5)
\nonumber \\
&\quad
+ \left(
-\frac{7539856966909}{738112500}
- \frac{2990809136}{273375}\zeta(3)
+ \frac{1254952}{405}\zeta(4)
+ \frac{2579312}{405}\zeta(5)
\right)N_f
\nonumber \\
&\quad
+ \left(
\frac{13542385424}{61509375}
+ \frac{17168}{45}\zeta(3)
- \frac{5024}{27}\zeta(4)
\right)N_f^2
+ \left(
-\frac{17813699}{8201250}
+ \frac{10048}{1215}\zeta(3)
\right)N_f^3, \\
\gamma_{{\rm VDDD},1}^{\mathrm{RI}'}
&= \frac{12383774}{30375}
- \frac{18466}{675}N_f + \frac{16}{\sum_\mu p^2/p_\mu^2} r_{{\rm VDDD},1}, \\
\gamma_{{\rm VDDD},2}^{\mathrm{RI}'}
&= \frac{1531697925821}{82012500}
- \frac{897704}{405}\zeta(3)
+ \left(
-\frac{3391730011}{1366875}
- \frac{6752}{135}\zeta(3)
\right)N_f
+ \frac{5838226}{91125}N_f^2
\nonumber \\
& \quad + \frac{16}{\sum_\mu p^2/p_\mu^2} r_{{\rm VDDD},2}, \\
\gamma_{{\rm VDDD},3}^{\mathrm{RI}'}
&= \frac{36856806095674111}{35429400000}
- \frac{836461717442}{4100625}\zeta(3)
+ \frac{143105822}{3645}\zeta(5)
\nonumber \\
&\quad
+ \left(
-\frac{6103983117441629}{29524500000}
+ \frac{1678175072}{273375}\zeta(3)
+ \frac{394084}{135}\zeta(5)
\right)N_f
\nonumber \\
&\quad
+ \left(
\frac{698462539903}{61509375}
+ \frac{169072}{405}\zeta(3)
\right)N_f^2
- \frac{698495693}{4100625}N_f^3 + \frac{16}{\sum_\mu p^2/p_\mu^2} r_{{\rm VDDD},3}, \\
r_{{\rm VDDD},1}
&= \frac{44}{5} - \frac{8}{15}N_f,
\\
r_{{\rm VDDD},2}
&= \frac{22664999}{3375}
+ \frac{45632}{45}\zeta(3)
+ \left(
-\frac{7773983}{6750}
- \frac{5152}{45}\zeta(3)
\right)N_f
+ \frac{3332}{75}N_f^2,
\\
r_{{\rm VDDD},3}
&= \frac{12607614183851}{97200000}
- \frac{1143851501}{30375}\zeta(3)
+ \frac{576884}{27}\zeta(5)
\nonumber \\
&\quad
+ \left(
-\frac{3198925062481}{145800000}
+ \frac{123562982}{91125}\zeta(3)
- \frac{104888}{81}\zeta(5)
\right)N_f
\nonumber \\
&\quad
+ \left(
\frac{360258397}{303750}
+ \frac{12080}{243}\zeta(3)
\right)N_f^2
- \frac{207836}{10125}N_f^3.
\end{align}
\end{widetext}

The quantity $\Delta Z_{X}^{{\rm RGI}, S}(\mu)$ is defined as~\cite{Gockeler:2010yr}
\begin{align}
    \Delta Z_{X}^{{\rm RGI}, S}(\mu) =& (2 \beta_0 \, a^S (\mu))^{-\gamma_{X,0}/(2 \beta_0)} \, \times \nonumber \\
    & \hspace{-1cm}\exp \left[ \int_0^{g^S(\mu)} dg' \left(\frac{\gamma^S_{X} (g')}{\beta^S (g')} + \frac{\gamma_{X,0}}{\beta_0 \, g'}\right) \right].
\end{align}
Expanding over $g^S$, we end up with
\begin{align}
    &\Delta Z_{X}^{{\rm RGI}, S} = (2 \beta_0 \, a^S)^{-\gamma_{X,0}/(2 \beta_0)} \, \times \nonumber \\
    & \qquad [1 + d^S_{X,1} a^S + d^S_{X,2} (a^S)^2 + d^S_{X,3} (a^S)^3 + \mathcal{O}((a^S)^4)],
\end{align}
where
\begin{widetext}
\begin{align}
d^S_{X,1} &=
\frac{\beta_1\gamma_{X,0}-\beta_0\gamma^S_{X,1}}
{2\beta_0^2},
\\
d^S_{X,2} &=
\frac{
-2\beta_0^3\gamma^S_{X,2}
+\beta_0^2\left[
\gamma^S_{X,1}
\left(2\beta_1+\gamma^S_{X,1}\right)
+2\beta_2^S\gamma_{X,0}
\right]
-2\beta_0\beta_1\gamma_{X,0}
\left(\beta_1+\gamma^S_{X,1}\right)
+\beta_1^2\left(\gamma_{X,0}\right)^2
}
{8\beta_0^4},
\\
d^S_{X,3} &=
\frac{1}{48\beta_0^6}
\Bigg[
\beta_1^3\left(\gamma_{X,0}\right)^3
-3\beta_0\beta_1^2\left(\gamma_{X,0}\right)^2
\left(2\beta_1+\gamma^S_{X,1}\right)
+\beta_0^2\beta_1\gamma_{X,0}
\left(
8\beta_1^2
+6\beta_2^S\gamma_{X,0}
+12\beta_1\gamma^S_{X,1}
+3\left(\gamma^S_{X,1}\right)^2
\right)
\\
& \qquad \qquad
+\beta_0^4
\left(
8\beta_3^S\gamma_{X,0}
+8\beta_2^S\gamma^S_{X,1}
+8\beta_1\gamma^S_{X,2}
+6\gamma^S_{X,1}\gamma^S_{X,2}
\right)
\nonumber \\
&\qquad \qquad
-\beta_0^3
\Big(
16\beta_1\beta_2^S\gamma_{X,0}
+8\beta_1^2\gamma^S_{X,1}
+6\beta_2^S\gamma_{X,0}\gamma^S_{X,1}
+\left(\gamma^S_{X,1}\right)^3
+6\beta_1
\left[
\left(\gamma^S_{X,1}\right)^2
+\gamma_{X,0}\gamma^S_{X,2}
\right]
\Big)
-8\beta_0^5\gamma^S_{X,3}
\Bigg].
\end{align}
\end{widetext}
The coupling constant $g^S (\mu)$ is determined by solving Eq.~\eqref{beta} up to four loops~\cite{Alekseev:2002zn}, as follows:
\begin{widetext}
\begin{equation}
a^S(\mu) = \frac{1}{\beta_0 L}
\Bigg[
1- \frac{\beta_1}{\beta_0^2}\frac{\ln L}{L}
+ \frac{
-\beta_1^2
+\beta_0\beta_2^S
+\beta_1^2(\ln L-1)\ln L
}{
\beta_0^4 L^2
}
+ \frac{
-\beta_1^3
+\beta_0^2\beta_3^S
+\beta_1\ln L
\left(
4\beta_1^2
-6\beta_0\beta_2^S
+\beta_1^2(5-2\ln L)\ln L
\right)
}{
2\beta_0^6 L^3
}
\Bigg],
\end{equation}
\end{widetext}
where $L = \ln(\mu^2/\Lambda_{\rm QCD}^2)$ and $\Lambda_{\rm QCD} = 295(10)~{\rm MeV}$ is the value quoted by FLAG for $N_f=4$ flavors~\cite{FlavourLatticeAveragingGroupFLAG:2024oxs}.

\end{document}